\documentclass[final,5p,times,reprint,authoryear]{elsarticle}

\usepackage{amssymb}
\usepackage{textcomp,gensymb}
\usepackage{setspace}
\usepackage{amsthm}
\usepackage{breqn}
\usepackage{amsmath}
\usepackage{float}
\usepackage{soul}
\usepackage{array}
\usepackage{booktabs}
\usepackage{svg}
\usepackage[ruled,vlined]{algorithm2e}
\SetKw{KwBy}{by}
\usepackage{hyperref}

\hypersetup{
    colorlinks = true,
    urlcolor   = red,
    citecolor  = red,
 }
 \hypersetup{breaklinks=true}

\journal{Chemical Engineering Science}
\newcommand{\beginsupplement}{%
        \setcounter{table}{0}
        \renewcommand{\thetable}{S\arabic{table}}%
        \setcounter{figure}{0}
        \renewcommand{\thefigure}{S\arabic{figure}}%
     }
\begin{document}
\begin{frontmatter}
\title{Comparison of empirical and particle force-based density segregation models}

\author[inst1]{Soniya Kumawat}
\author[inst2]{Vishnu Kumar Sahu}
\cortext[cor1]{Department of Chemical Engineering, Indian Institute of Technology Kanpur, Uttar Pradesh, 208016, India}
\author[inst1]{Anurag Tripathi \texorpdfstring{\corref{cor1}} {}}
\ead{anuragt@iitk.ac.in}
\affiliation[inst1]{organization={Department of Chemical Engineering},
            addressline={Indian Institute of Technology Kanpur}, 
            city={Uttar Pradesh},
            postcode={208016}, 
            country={India}}
\affiliation[inst2]{organization={Department of Chemical Engineering and Materials Science},
            addressline={University of Minnesota}, 
            city={Minneapolis},
            postcode={MN 55455}, 
            country={USA}}            
\begin{abstract}

The empirical and particle force-based models of granular segregation due to density differences among the species are compared in this work. Dependency of the empirical segregation parameters on the initial configuration, the observation time duration, inclination angle, and mixture composition are discussed in detail. The parameters obtained from empirical models are used to predict the steady-state concentration profiles for different density ratios and compositions. In addition, we utilize the predictions from the particle force-based segregation model and compare them with the predictions of the empirical segregation models. Our results show that the linear empirical segregation model predictions agree well with the simulation results for mixtures rich in light species where as quadratic empirical segregation model works better for mixtures rich in heavy species. Particle force-based segregation model, on the other hand, seems to be in very good agreement with the DEM simulation data across all mixture compositions.
\end{abstract}
\begin{highlights}
\item Measurement duration and initial configuration influence empirical parameter values
\item Linear model parameters are more sensitive compared to quadratic model parameters 
\item Empirical models show larger deviations compared to particle force based models
\end{highlights}
\begin{keyword}
Granular mixtures, Discrete Element Method, Density segregation, Empirical segregation model, Particle-force based segregation model
\end{keyword}
\end{frontmatter}
\newpage
\section{Introduction}

Granular mixtures often undergo segregation during their flow due to differences in the sizes and$/$or densities of the particles.
Variety of geometries such as chute  \citep{tripathi2013density, deng2018continuum, tripathi2021size}, vibrated bed \citep{YANG2006, TAI2010, jain2013effect}, hopper \citep{xiao2019continuum},  heap \citep{Baxter1998, schlick2016, fan2017segregation, Deng2020}, and rotating tumbler \citep{pereira2014segregation, schlick2015granular, Chen2016, deng2019modeling} have been used to understand and predict the segregation of granular mixtures.
Discrete element method simulations are commonly utilized to obtain various properties of the flowing mixture which are then used to propose segregation models, often in the form of correlations. 
These segregation models are typically used in the continuum framework by employing the convection-diffusion equation to predict the concentration of different species \textcolor{black}{in flowing granular mixtures}. 

One of the most popular approaches for proposing these segregation models involves empirically correlating the species segregation velocity to the flow properties of the mixture. \textcolor{black}{Early researchers investigating segregation of different size mixtures (\cite{savage1988particle, gray_chugunov_2006, gray_ancey_2011}) \textcolor{black}{assumed} the segregation velocity of the species to be linearly dependent on the concentration of other species.} 
Using the segregation model of \textcolor{black}{\cite{gray_chugunov_2006}, however, showed significant difference from the experimental concentration profiles for a binary mixture of different size grains in the study of \cite{Wiederseiner2011}.} The authors attributed these differences in the concentration profiles to possible dependence of the segregation velocity on the local shear rate.
Later studies (\cite{fan2014modelling, schlick2015modeling}) confirmed this local shear rate dependency of the segregation velocity and utilized the combined shear rate and concentration dependent segregation velocity to predict size segregation in heap flow.
 Both linear (\cite{fan2014modelling, schlick2015modeling, schlick2016, deng2018continuum, deng2019modeling, Deng2020,GAO2021}) as well as nonlinear (\cite{gajjar2014asymmetric, Vandervaart2015, jones2018asymmetric}) forms have been proposed to account for the concentration dependency. 
\textcolor{black}{A linear dependency of the segregation velocity has been found on the shear rate in all these studies.}

\textcolor{black}{This dependency on the shear rate as well as the species concentration has been also confirmed for mixtures with same size but different density particles}  (\cite{xiao2016modelling, fry2018effect, jones2018asymmetric, fry2019}). 
By performing DEM simulations of quasi 2D heap flows, \cite{xiao2016modelling} obtained the linear dependence of the segregation velocity $v_{seg,i}$ on the local concentration of other species $f_j$ and shear rate for binary mixtures differing in density. The authors proposed a correlation for the segregation velocity $v_{seg,i} = S_D \dot \gamma f_j$ as proposed by \cite{fan2014modelling} and \cite{schlick2015modeling} in the case of size segregation.
The segregation length scale $S_D$ in different density mixtures was found to be proportional to the diameter $d$ of the particles and \textcolor{black}{logarithm} of the density ratio $\rho$ and was given as $S_D = d C_D \ln{\rho}$. The value of the empirical parameter ($C_D$) was estimated by linearly fitting the $S_d$ vs. $\rho$ data on a semi-log plot. 
\cite{fry2018effect} modified \textcolor{black}{this} linear empirical segregation model to account for the effect of the confining pressure in addition to the local concentration and shear rate in case of plane shear flows. 
\textcolor{black}{In a followup study \citep{fry2019}, the authors showed that the model is able to predict segregation in plane shear flow geometry for different densities as well as different size mixtures.}

\cite{jones2018asymmetric} studied the segregation for a binary mixture due to density difference using DEM simulations in $2D$ bounded heap flow. The authors also investigated the size segregation of binary mixtures in the same geometry. Based on their findings, the authors revised the empirical model for the segregation velocity and proposed a quadratic dependence on the concentration along with the linear dependency on the shear rate. 
\cite{Duan2021} showed that the combined effect of size and density segregation in $2D$ quasi-heap flow can also be described well using the quadratic empirical model. The empirical parameters are found to be dependent on the size ratio as well as the density ratio. \cite{duan2022segregation} showed that this approach is also able to capture the segregation in a rotating tumbler. 

Another promising approach to predict segregation is referred to as the particle force-based approach. By considering the forces on a high density intruder settling in a flowing granular medium, \cite{tripathi2013density} proposed the particle force-based theory for predicting density segregation \textcolor{black}{of} binary mixtures in chute flow. 
In a recent study \citep{sahu_kumawat_agrawal_tripathi_2023}, the particle force-based theory is extended for multi-component granular mixtures of different densities. This model is successfully able to capture the segregation in binary and multi-density mixtures. 
A similar approach is utilized by \cite{tripathi2021size} to obtain the theoretical expression for the segregation velocity of large size particles and predict the concentration profiles in binary size mixtures for periodic chute flows. 

Both empirical as well as particle force-based approaches account for the effect of the species concentration and shear rate. However, the dependence on the species densities appears to be very different in these two approaches. While the particle force-based model suggests segregation velocity to be proportional to the density difference between the same size particles, the linear empirical model suggests that the density ratio of the species determines the segregation. A quick comparison of the particle force-based segregation velocity expression given by \cite{duan2020DenSegKTGF} indeed suggests that the segregation parameter for the linear empirical model is proportional to the density difference and inversely proportional to the product of pressure and effective friction coefficient (i.e., the shear stress). Given the fact that the shear stress in the case of chute flows depends on the species concentration and inclination angle, the dependence of the empirical segregation parameter on these two can't be ruled out. Further, the particle force-based theory suggests strong inter-coupling of rheology and segregation with each other during the flow. The flow kinematics for systems starting from different initial configurations may be significantly different due to such inter-coupling of rheology and segregation. Since the empirical segregation parameters are obtained from the kinematics data during the flow evolution, the dependency of these parameters on flow configuration also remains to be explored. 
In this study, we compare the empirical and particle force-based segregation models in an identical geometry. Specifically, we chose periodic chute flow of different density particles over a bumpy base and explore the effect of mixture composition, measurement time duration, density ratios at different inclination angles, and different initial configurations on the empirical segregation parameters. 

The organization of the paper is as follows.  Section \S \ref{Sec:Sim-methodology} describes the simulation methodology for DEM simulations. Details of the calculation of the empirical segregation parameters estimation 
along with the effect of the measurement time and initial configuration are discussed in \S \ref{sec:empirical-parameter}. \textcolor{black}{This detailed analysis helps us identify the limitations of the empirical models and determine the best way to estimate the appropriate values of the empirical parameters.} 
The theory utilizing the two empirical models to predict the concentration profile in the periodic chute is presented in \S \ref{sec:Comp_Emp_and_Ptcle}. In addition, the predicted profiles are compared with the DEM simulation and particle force-based theory predictions in this section. \textcolor{black}{The conclusions are presented in section \S \ref{sec:conclusion}.}

\section{Simulation methodology}
\label{Sec:Sim-methodology}
\begin{figure}[htbp]
    \centering
    \includegraphics[scale=0.35]{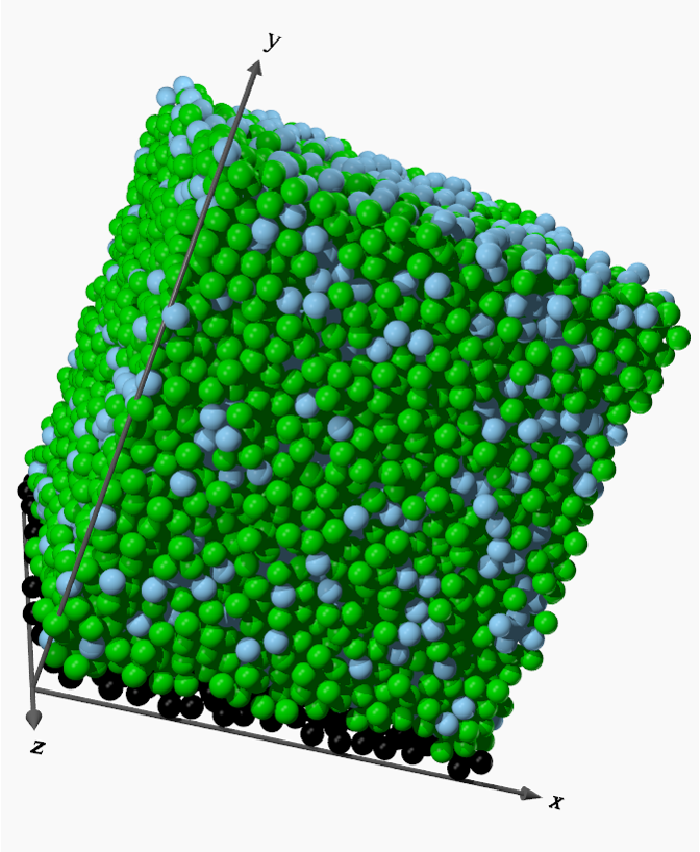}\put(-110,135){(a)} 
        \includegraphics[scale=0.35]{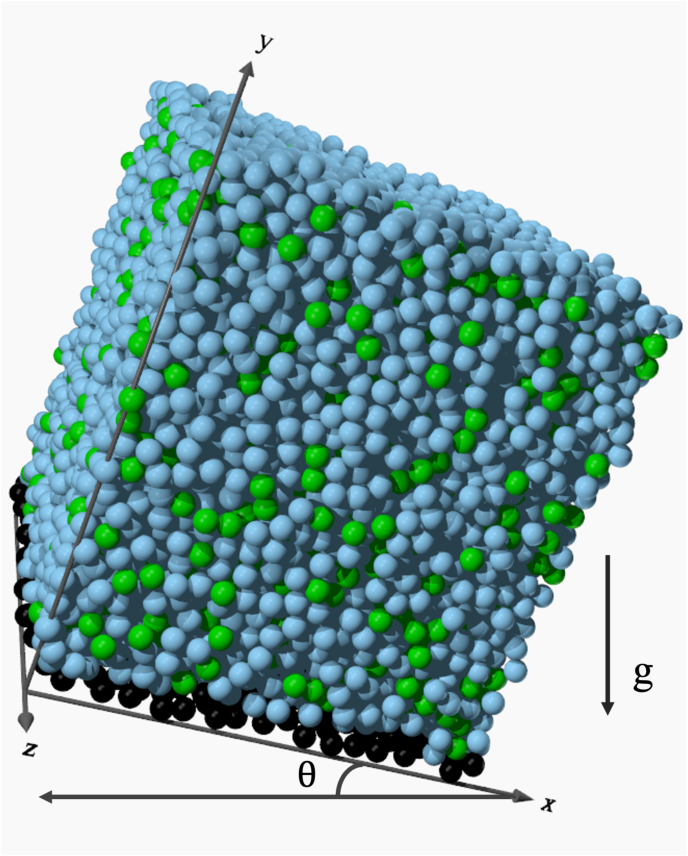}\put(-110,135){(b)}
    \caption{DEM simulation snapshots for binary granular mixture having (a) $20\%$ and (b) $80\%$ light particles  with density ratio $\rho = 3.0$ flowing over an inclined plane at an inclination angle of $\theta$. Green and blue particles represent the heavy and light particles, respectively. Black particles represent the rough bumpy base of thickness $1.2d$.}
    \label{fig:Sim-methodoloy}
\end{figure}

We simulate granular mixtures of different density particles, modelled as slightly inelastic and frictional spheres, flowing over a bumpy inclined surface. 
At the beginning of the simulation, the particles are positioned in such a way that their centers form a cubic lattice, ensuring that there is no contact between any pair of particles. Additionally, these particles are assigned random initial velocities. First, we settle the particles at a zero inclination angle and then increase the inclination to the desired angle. Figure~\ref{fig:Sim-methodoloy} shows the uniformly mixed arrangement of high-density (green) particles of mass $m_{H}$ and the light density (blue) particles of mass $m_{L}$ \textcolor{black}{for two different mixture compositions}. 
The bumpy chute base, shown as black spheres in figure~\ref{fig:Sim-methodoloy}, is formed by taking a slice of $1.2d$ thickness from a randomly arranged, densely packed arrangement of spheres of the diameter of $d$. 
\textcolor{black}{The simulation box has equal length and width of $20d$ in the flow ($x$) and the neutral ($z$) directions.} 
The height of the simulation box in the $y$-direction is chosen to be sufficiently large in order to make sure that the particles never reach the top surface to effectively mimic free surface flow. 
A total of $10000$ flowing particles are simulated within the system so that the typical flowing layer height $H\approx 25d$. 
The simulations are performed for binary as well as ternary density mixtures over a wide range of compositions, and inclination angles. \textcolor{black}{We consider three different initial configurations for density segregation: light near base, heavy near base, and well-mixed configuration.} 

The flow and segregation behavior of the mixtures is simulated using the Discrete Element Method (DEM) simulation. 
Following \cite{tripathi2011numerical}, we use linear spring dashpot contact force model, and the Coulomb's friction law is used to limit the maximum permissible tangential force. The equations of motion are numerically integrated using a sufficiently small time step. Both gravity as well as inter-particle contact forces acting on the particles are accounted for to obtain the positions and velocities of every particle.
All the particle properties used in the DEM simulations are identical to those used in \cite{tripathi2011numerical} and \cite{sahu_kumawat_agrawal_tripathi_2023}. Using the positions and velocities of the particles, all the relevant quantities of interest are calculated. Due to the periodic boundaries in the $x$ and $z$ direction, properties are obtained across different strips in the $y$ direction. The sampling volume is taken as a strip of thickness $d$ spanning the entire area of the simulation box in the $x-z$ plane. In order to get a good statistical average, the properties are averaged over a total of $20$ time units obtained at regular intervals of $0.1$ time unit.

Packing fraction of $j^{th}$ species in the mixture is calculated as $\phi_j = (1/V)\sum_i \phi_{j,i}$, where $\phi_{j,i}$ is the volume fraction of $i^{th}$ particle of the $j^{th}$ species in the sampling volume $V$. The total packing fraction of the mixture is obtained as $\phi = \sum_j \phi_j$. The concentration of $j^{th}$ species at any $y$ location is obtained as 
\begin{equation}
    f_j =  \frac{\phi_j}{\phi}.
\end{equation}
Species velocity is calculated as, 
\begin{equation}
   \vec v_j =\frac {\sum \vec c_{j, i} \phi_{j, i} } {\sum \phi_{j, i}},
\end{equation}
where $ \vec c_{j, i }$ is the instantaneous velocity of particle $i$ belonging to species $j$. The average velocity of all the $N$ particles partially or completely situated in the sampling volume $V$ is calculated as $\vec{v}=\sum_{k=1}^{N} \phi_k \vec {c}_{k}$ where $\phi_k$ is the volume fraction of particle $k$ in the sampling volume and $\vec {c}_{k}$ is the instantaneous velocity of particle. The shear rate at each $y$ location is calculated by numerically differentiating the velocity profile using forward difference method. Following \cite{fan2014modelling}, \cite{xiao2016modelling}, and \cite{jones2018asymmetric}, the segregation velocity of $j^{th}$ species $\vec v_{seg,j}$ is defined as the local velocity of species $j$ with respect to mean flow $\vec v_{mix}$ in the cases where contribution due to diffusional velocity is neglected, i.e., 
\begin{equation}
    \vec v_{seg,j} =  \vec v_j -  \vec v_{mix}.
    \label{eq:segVel_vp-v}
\end{equation}
\textcolor{black}{Data for the segregation velocity is obtained for each strip of thickness $1d$ across the entire flowing layer and is averaged over an interval of $100$ time units.} Steady-state is characterized by a constant kinetic energy of the system and a constant height of the centre of masses of all the species in the mixture. 

\section{Calculation of empirical segregation parameter}
\label{sec:empirical-parameter}
\begin{figure*}[!htb]
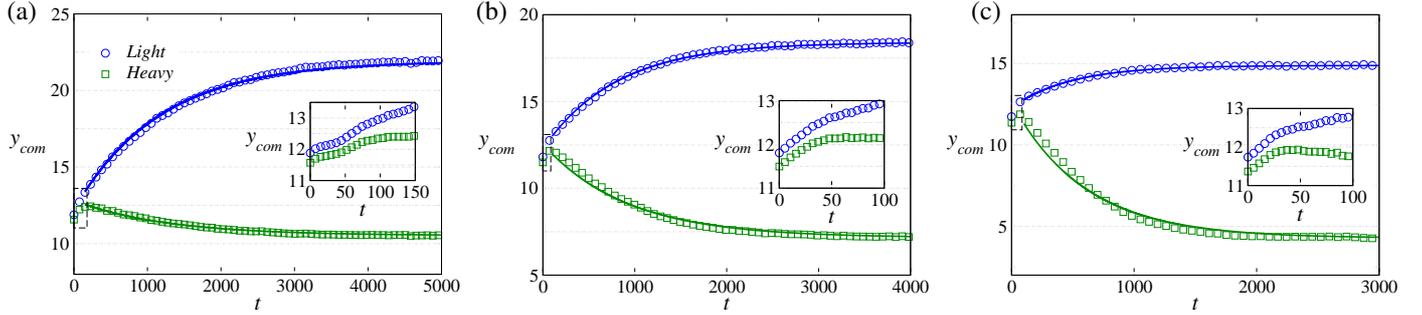

    \centering
    \includegraphics[scale=0.32]{Figures/Fig2a.eps}\put(-170,110){(a)}\hfill
     \includegraphics[scale=0.32]{Figures/Fig2b.eps}\put(-170,110){(b)}\hfill
     \includegraphics[scale=0.32]{Figures/Fig2c.eps}\put(-160,110){(c)}\hfill
    \caption{Centre of mass variation of light (blue circles) and heavy (green squares) species with time for a binary mixture having density ratio $\rho = 3.0$ starting from an initially well-mixed configuration. The composition of the light species in the mixture is (a) $f^T_{L} = 0.20$ , (b) $f^T_{L} = 0.50$, and (c) $f^T_{L} = 0.80$. Symbols represent the DEM data and the lines are the exponential fits to the data.}
    \label{fig:ycom}
\end{figure*}

A binary granular mixture flowing over an inclined surface is known to segregate in regions rich in one component. 
Previous studies (see \cite{fan2014modelling,schlick2015modeling,xiao2016modelling,jones2018asymmetric,Duan2021}) use the empirical approach to predict the segregation due to size or density differences in free surface flows such as heap and rotating tumbler. 
All of these approaches utilize the flow and kinematics data starting from an initially well-mixed configuration for obtaining segregation parameters. 
Since this parameter determination utilizes the data during the flow evolution, the possibility of empirical parameters being dependent on the duration of observation can't be ignored.  
Hence, we performed DEM simulations and explored the effect of the observation time duration on the empirical parameter calculation. We consider a periodic chute flow simulation of a binary mixture having two different density particles. \textcolor{black}{The usage of this setup provides a much better control over the inclination angle (and the shear rate) compared to other setups and facilitates investigation of the effect of the measurement time duration on empirical segregation parameters.} 
\textcolor{black}{Using this simple setup, we estimate the segregation time scale of the species using the data of centre of mass variation with the time.} 
We first, discuss the duration of segregation for different compositions of binary density mixtures.
Figure~\ref{fig:ycom} shows the variation of the centre of mass ($y_{com}$) with time for three different mixtures. 
Figure \ref{fig:ycom}a shows the data for a mixture having $20\%$ light species ($f^T_{L} = 0.20$) whereas figures \ref{fig:ycom}b and \ref{fig:ycom}c show the data for $f^T_{L} = 0.50$ and $f^T_{L} = 0. 80$ respectively. The data presented corresponds to a binary mixture with a density ratio of $\rho = 3.0$ flowing over an inclined plane at inclination $\theta = 25\degree$. 
Following the studies mentioned in the beginning of this section, we start the simulations from an initial condition where the light and heavy particles are uniformly mixed.
Hence, the centre of mass position of both the species at time $t = 0$ is nearly identical for all three cases. 
The particles are settled under the influence of gravity at zero inclination angle, resulting in a packing fraction close to a randomly closed-packed configuration. Shearing of such a closed-packed layer undergoes dilation as it starts to flow. Such a dilation at early times is observable in the inset of figures \ref{fig:ycom}a, \ref{fig:ycom}b, and \ref{fig:ycom}c. Note that both heavy and light species exhibit a rise in the centre of mass position \textcolor{black}{in the beginning} due to the dilation of the layer. After an initial period of dilation, the centre of mass position of heavy species (green squares) decreases while that of light species (blue circles) increases with time. The $y_{com}$ position of both the species becomes constant at larger times, indicating that steady-state segregation is achieved.

Ignoring the dilation for the initial times, the subsequent centre of mass variation can be captured very well using an exponential function. 
The following mathematical expression is used to fit the data shown in figures~\ref{fig:ycom}a-\ref{fig:ycom}c, 
\begin{equation}
    y_{com}(t) = y^0_{com} + (y^{\infty}_{com} - y^0_{com})\left(1-e^{-(t-t_0)/\tau}\right). 
    \label{eq:exp-fitting}
\end{equation}
In this expression, $y^0_{com}$ and $y^{\infty}_{com}$ are the initial and final centre of mass positions of the species. 
$\tau$ is the time scale that characterizes the segregation phenomena and $t_0$ is the dilation time beyond which $y_{com}$ of heavy species starts to decrease. 
A similar expression has been utilized by \cite{timescale2014} to calculate the segregation time scale. 
The exponential curve (shown using solid lines in figure~\ref{fig:ycom}) describes the data reasonably well for both the species for all three mixture compositions. 
The segregation time scale $\tau$ obtained by fitting equation \ref{eq:exp-fitting} to the species centre of mass profiles is reported in table~\ref{tab:time-scale}. Note that the segregation time scale is found to be different for the two species and varies with the composition of the mixture.

\begin{table}[h]
\centering
\renewcommand{\arraystretch}{1.5}
\begin{tabular}{p{3.2cm}ccc}
\hline
Segregation time scale & \textbf{$f^T_{L} = 0.2$} & \textbf{$f^T_{L} = 0.5$} & \textbf{$f^T_{L} = 0.8$} \\
\hline
Heavy species ($\tau_H$) & 1157 & 799 & 478 \\
Light species ($\tau_L$) & 1237 & 916 & 599 \\
\hline
\end{tabular}
 \caption{Segregation time scale for heavy and light species for three different mixture compositions shown in fig~\ref{fig:ycom}}
    \label{tab:time-scale}
\end{table}

Table~\ref{tab:time-scale} also shows that the time scale for both the species decreases with an increase in the composition of the light particles. Additionally, the time scale of the light species is found to be approximately $80$ to $120$ time units larger than that of heavy species.
The average value of the time scales of the two species i.e., $\tau = (\tau_H + \tau_L)/2$ is taken as the mixture segregation time scale for further analysis.

\subsection{Linear empirical models}
\label{subsec:linearEmp}
\begin{figure*}[!htb]
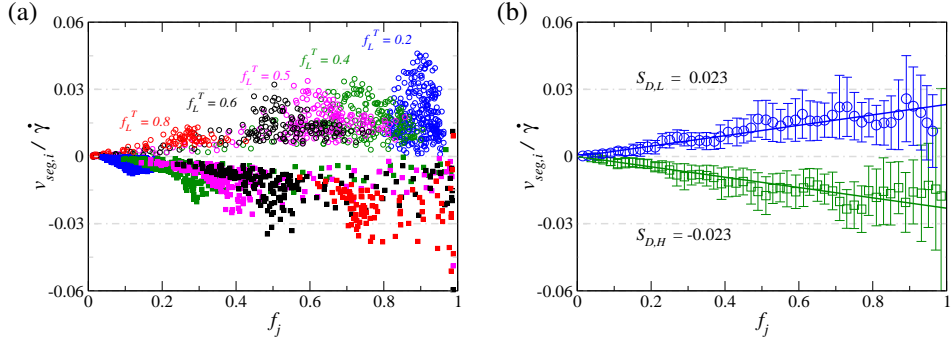

    \centering
    \includegraphics[scale=0.3]{Figures/Fig3a.eps}\put(-170,120){(a)}\quad \quad
    \includegraphics[scale=0.3]{Figures/Fig3b.eps}\put(-170,120){(b)}\quad \quad
    \caption{(a) DEM simulation data of shear rate scaled segregation velocity for light (circles) and heavy (squares) species of a binary mixture having density ratio $\rho = 3.0$ flowing over chute at inclination angle $\theta = 25\degree$. Different colours correspond to different mixture compositions of light species in the range $f^T_{L} = 0.2$ to $f^T_{L} = 0.8$. (b) DEM simulation data for all the compositions averaged over every $\Delta f_i = 0.02$. Solid lines represent the linear fit to the averaged data.}
    \vspace{1.7cm}
    \label{fig:density-Raw-data}
\end{figure*}

\begin{figure*}[!htb]
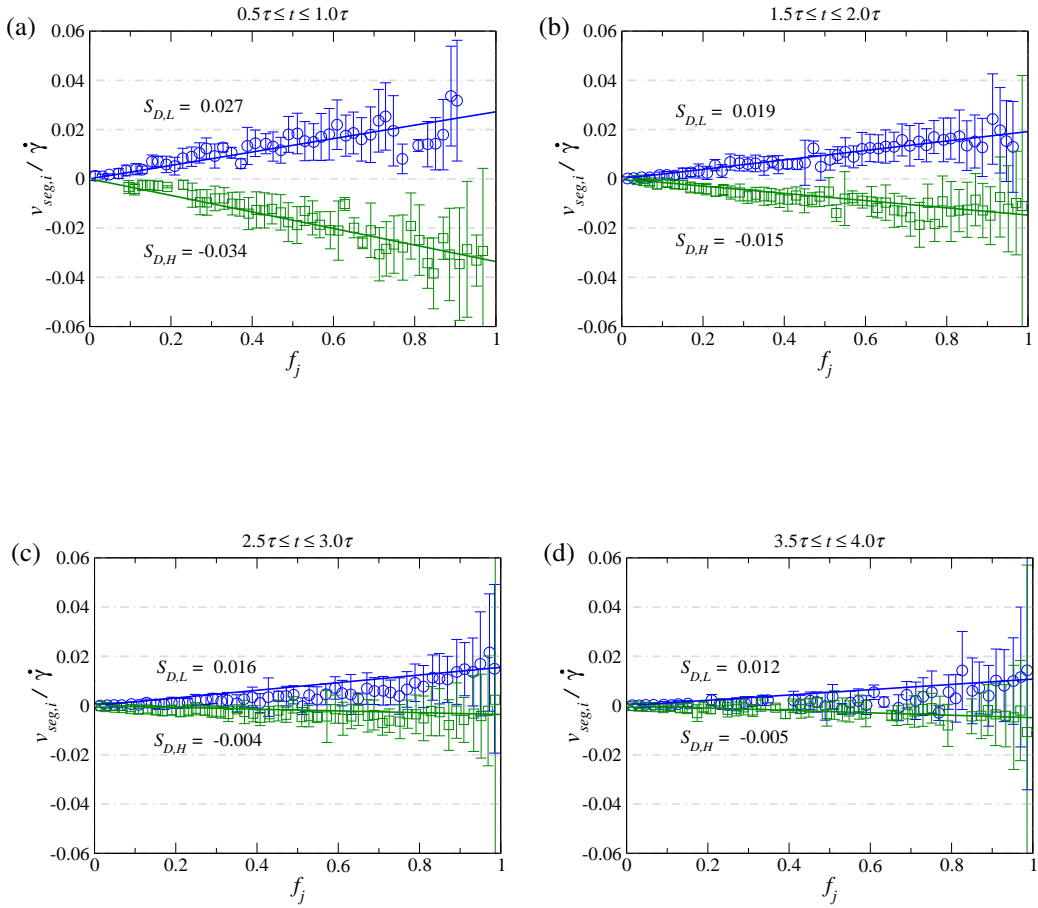

    \centering
   \quad \quad \quad \quad  \quad \quad \quad \includegraphics[scale=0.33]{Figures/Fig4a.eps}\put(-185,130){(a)}\quad \quad 
    \includegraphics[scale=0.33]{Figures/Fig4b.eps}\put(-185,130){(b)}\hfill
    \vspace{2cm}
    \includegraphics[scale=0.33]{Figures/Fig4c.eps}\put(-185,130){(c)}\quad \quad
     \includegraphics[scale=0.33]{Figures/Fig4d.eps}\put(-185,130){(d)}\hfill
    \caption{Dependency of the segregation parameter on the observation time duration (a) $0.5\tau \leq t \leq 1.0\tau$, (b) $1.5\tau \leq t \leq 2.0\tau$, (c) $2.5\tau \leq t \leq 3.0\tau$ and (d) $3.5\tau \leq t \leq 4.0\tau$. Blue symbols represent the light species, and green symbol represents the heavy species. Solid lines represent the linear fit $v_{seg,i} / \dot{\gamma} = S_D f_j$.  Data reported for inclination angle $\theta=25^\circ$  and density ratio $\rho=3.0$.}
    \label{fig:timerange}
\end{figure*}

Following \cite{fan2014modelling}, \cite{schlick2015modeling} \& \cite{xiao2016modelling}, we define the segregation velocity as the species velocity relative to the mixture velocity using equation \ref{eq:segVel_vp-v}. 
Using this definition, these studies have shown that the species segregation velocity scaled by the local shear rate i.e, $v_{seg,i}/\dot \gamma$, varies linearly with other species concentration $f_j$ for bounded heap flows and rotating cylinders. Here, we report the data for the segregation velocity in periodic chute flow obtained from DEM simulations in figure~\ref{fig:density-Raw-data}.
These data correspond to a binary mixture density ratio $\rho=3.0$ flowing at an inclination angle of $\theta = 25\degree$. We consider the data for an observation duration of approximately two-time scales ($t_0 \leq t \leq 2\tau$). The data during the initial dilation period $t < t_0$ is not included. Following \cite{xiao2016modelling}, the flow is started from an initially mixed configuration. The light species tends to rise towards the free surface, resulting in a positive segregation velocity. Conversely, heavy species move towards the base, leading to a negative segregation velocity.

Figure \ref{fig:density-Raw-data}a shows the variation of the segregation velocity scaled by the shear rate $v_{seg,i}/\dot \gamma$ with the concentration of other species $f_j$. 
Different coloured symbols represent the local segregation velocity for different mixture compositions. Total five different mixtures compositions with $f^T_L = 0.2, 0.4, 0.5, 0.6$, and $0.8$ have been considered to ensure the full range of $f_j$ from $0$ to $1$ in figure \ref{fig:density-Raw-data}.
The large scatter in the data points of figure~\ref{fig:density-Raw-data}a is substantially reduced by averaging it over bins of width $\Delta f_i = 0.02$ spanning over the entire concentration range.
This data, shown in figure \ref{fig:density-Raw-data}b, can be described very well using a straight line.
The slope of the fitted line for the heavy species is $S_{D, H} = -0.023$ while that for the light species is $S_{D, L} = 0.023$.
Note that the magnitude of slopes (termed as the segregation or percolation length scale in previous studies) for both species is identical. 
These results indicate that the segregation velocity scaling proposed by \cite{xiao2016modelling} for the density segregation holds true in the case of the chute \textcolor{black}{flow} as well. Thus, the segregation velocity of species $i$ can be described using the following relation,
\begin{equation}
\centering
    v_{seg,i} = S_{D}(i,j) \dot \gamma f_j.
    \label{eq:vp_density}
\end{equation}

Recall that the data reported in figure~\ref{fig:density-Raw-data} correspond to an observation duration of approximately two-time scales. We now investigate the effect of the observation time on the empirical parameters. We first report the variation of the scaled segregation velocity with time. To explore this effect, we collect the scaled segregation velocity data for different time intervals. \textcolor{black}{The data for the initial duration of layer dilation are discarded.}
Figure~\ref{fig:timerange} illustrates the shear rate scaled segregation velocity for both the species for four distinct time intervals, each with a duration of $0.5\tau$. As before, the data are presented for a binary mixture with   $\rho=3.0$ at $25^\circ$ starting from the mixed configuration. The segregation velocity scaled by the shear rate as a function of concentration is shown for $0.5\tau \leq t \leq 1.0\tau$ in figure \ref{fig:timerange}a, for $1.5\tau \leq t \leq 2.0\tau$ in figure \ref{fig:timerange}b, for $2.5\tau \leq t \leq 3.0\tau$ in figure \ref{fig:timerange}c, and for $3.5\tau \leq t \leq 4.0\tau$ in figure \ref{fig:timerange}d. 
Symbols represent the DEM data, and lines represent the linear fit to data. The slope of the fitted lines is mentioned for each time interval.
The shear rate-scaled segregation velocities for both the species exhibit higher values within the time range of $0.5\tau \leq t \leq 1.0\tau$. However, these scaled velocities decrease as time progresses. Notably, the values of $v_{seg,i}/\dot\gamma$ for both the species are significantly low during the time interval of $2.5\tau \leq t \leq 3.0\tau$ \& $3.5\tau \leq t \leq 4.0\tau$. We find that the linear fit (equation~\ref{eq:vp_density}) accurately describes the data for all the time intervals. However, the empirical parameters estimated from the linear fit to DEM data differ for different time intervals.
\textcolor{black}{The decreasing slope of fitted lines in figure \ref{fig:timerange} confirms} that the species segregation velocity keeps decreasing with time and becomes nearly zero at steady state. \textcolor{black}{This is in contradiction with the empirical models for segregation velocity. The linear empirical relation of equation \ref{eq:vp_density} states that segregation velocity depends only on the shear rate $\dot \gamma$ and concentration of other species $f_j$.} At steady state, the shear rate has a finite value, and except for the limiting cases of $f_i = 0$ \& $1$, $f_j$ is also non-zero. Therefore, according to equation \ref{eq:vp_density} the segregation velocity should have a finite non-zero value at steady state. \textcolor{black}{We consider this issue again later while discussing the role of diffusional velocity in estimation of segregation velocity using DEM.}  The time-dependent slope of scaled percolation velocity suggests that a time-averaged slope may be more appropriate for usage in equation \ref{eq:vp_density}.
Hence we consider data from $t_0 \leq t \leq t_f$, where $t_f$ ranges from $1\tau$ to $2.5\tau$. 
For a time duration larger than $ 2.5 \tau$, the segregation velocity becomes \textcolor{black}{very small and remains} practically indistinguishable from zero as the segregation process is around $95\%$ complete. We find that even after averaging the data over various observation time frames, substantial differences remain noticeable and the value of $S_D$ varies substantially for different values of $t_f$. 
Table~\ref{tab:seg-length} reports the absolute value of the segregation length scale empirically obtained from the DEM simulations for four different observation times. The segregation parameter has a negative value for heavy species and a positive value for light species.

\begin{table}[h]
\centering
\renewcommand{\arraystretch}{1.5}
\begin{tabular}{p{3.2cm}cccc}
\hline
Segregation parameter & \textbf{$\tau$} & \textbf{$1.5 \tau$} & \textbf{$2 \tau$} & \textbf{$2.5 \tau$}\\
\hline
Heavy $|S_{D,H}|$ & 0.031 & 0.027 & 0.023& 0.020 \\
Light $|S_{D,L}|$ & 0.027 & 0.025 & 0.023 & 0.022 \\
Average ($S_{D}$) & 0.029 & 0.026 & 0.023& 0.021 \\
\hline
\end{tabular}
 \caption{Segregation length scale for heavy and light species for different observation time duration $t_0 \leq t \leq t_f$ for different values of $t_f$ in the range $\tau$ to 2.5$\tau$. Average value of the segregation parameter is calculated as $S_D=[|S_{D, L}|+|S_{D, H}|]/2$.}
    \label{tab:seg-length}
\end{table} 
While the magnitude of segregation parameters of the two species are found to be reasonably close to each other, the values of the parameter seem to decrease as observation time increases. This decrease in the slopes is attributed to the fact that more data near the steady state are considered in the case of the higher value of $t_f$.
It appears that a time duration of twice the segregation time scale is appropriate as it provides a long enough time window to capture most of the finite segregation velocity data without capturing the near-zero segregation velocity close to the steady state. \textcolor{black}{Hence, we use the segregation parameter obtained over twice of segregation time scale ($2\tau$) in the rest of the paper for the linear empirical model.}
\textcolor{black}{Using the data over $2\tau$,} we report the segregation length scale for a range of density ratios $1.5 \leq \rho \leq 10$. 
\begin{figure}[!htbp]
    \centering
    \includegraphics[scale=0.40]{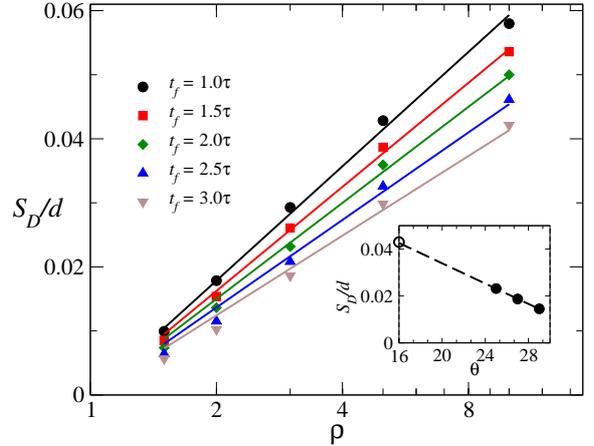}
    \caption{Variation of non-dimensional segregation parameter $S_D/d$ with density ratio $\rho = \rho_H/\rho_L$ on a semi-log plot for binary mixture at $\theta = 25\degree$ for different $t_f$.}
    \label{fig:Sd-vs-lnrho}
\end{figure}
Figure \ref{fig:Sd-vs-lnrho} shows that $S_D/d$ varies linearly with the density ratio on the semi-logarithm scale for all five different time durations considered in this study. The data presented in this figure corresponds to a binary mixture flowing at an inclination angle of $25^\circ$ starting from the mixed configuration. The data can be described using the relation $S_D(i,j)/d = C_D \ln (\rho_j/\rho_i)$ where $C_D$ is an empirical parameter. This relation is identical to that reported by \cite{xiao2016modelling}. 
\textcolor{black}{However, the time duration for segregation velocity measurement has not been mentioned by the authors.}
Our results show that the value of empirical parameter $C_D$ \textcolor{black}{(i.e the slope of fitted lines in figure~\ref{fig:Sd-vs-lnrho})} decreases by almost $23 \%$ by increasing the time duration from $1\tau$ to $3 \tau$. 
Figure \ref{fig:Sd-vs-lnrho} inset shows the data for $S_D/d$ variation at different inclination $\theta$ for a binary mixture of density ratio $\rho = 3.0$. Evidently, the $S_D/d$ value increases with decreasing the inclination angle $\theta$. The dashed line shows the extrapolated line for lower inclination angles. 
\cite{xiao2016modelling} reported that the segregation parameter $S_D$ is $0.15$ mm for the binary mixture having identical size particles of diameter $d = 3 $ mm and density ratio $\rho = 3.0$ flowing at inclination angle $\theta = 16\degree$ \textcolor{black}{(Figure 7 of their study).} 
Notably, the dimensionless value of $S_D = 0.05$ is found to be close to our extrapolated value of $0.043$.
\textcolor{black}{Based on their results,} we use the value of $C_D$ corresponding to a measurement time of $2 \tau$ while using the linear empirical segregation model and predict the theoretical concentration profile for binary as well as multi-component mixture flowing over a chute in section \S \ref{sec:Comp_Emp_and_Ptcle}.

\subsection{Quadratic empirical model}
\label{subsec:QuadraticEmp}
Linear empirical models (\cite{fan2014modelling}, \cite{schlick2015modeling}, and \cite{xiao2016modelling}) describing the segregation due to either size differences or density differences utilize equation~\ref{eq:segVel_vp-v} to measure the segregation velocity from DEM simulations.
A recent study by \cite{Duan2021} dealing with the combined size and density segregation suggests that this method of measurement of the segregation velocity using DEM simulations is not appropriate.
The authors revise the method to calculate segregation velocity and incorporate the presence of the diffusional velocity using the following expression
\begin{equation}
    v_{seg,i} = v_i - v_{mix} + \frac{1}{f_i} D\dfrac{\partial f_i}{\partial y}.
    \label{eq:duan_Vseg}
\end{equation}

\begin{figure*}[!htb]
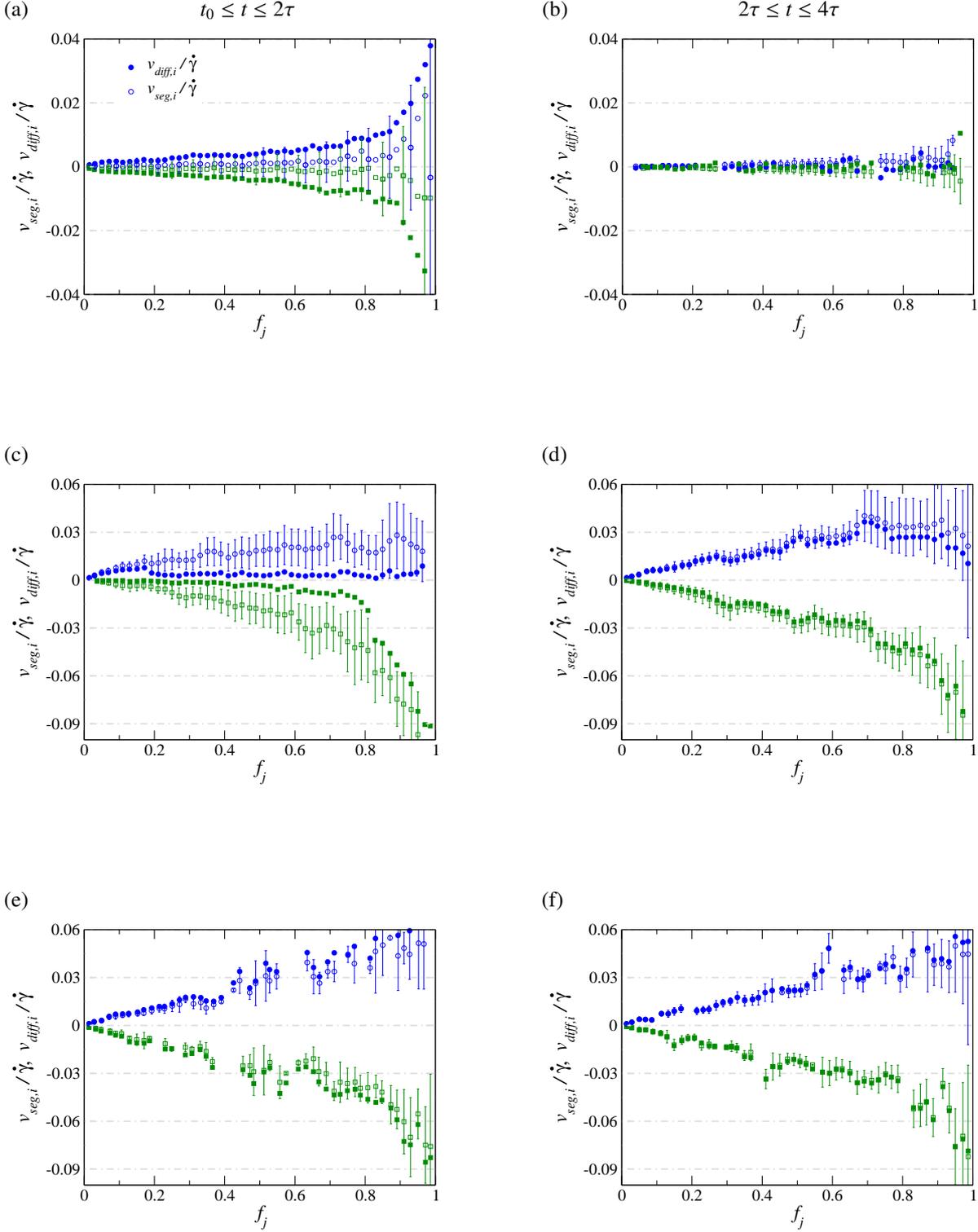

\includegraphics[scale=0.35]{Figures/Fig6a.eps}\put(-200,150){(a)}\put(-110,150){$t_0 \leq t \leq 2\tau$} \quad \quad \quad \quad \quad 
   \includegraphics[scale=0.35]{Figures/Fig6b.eps}\put(-200,150){(b)}\put(-110,150){$2\tau \leq t \leq 4 \tau$}\hfill
   \vspace{1.7cm}

 \includegraphics[scale=0.35]{Figures/Fig6c.eps}\put(-200,150){(c)}\quad \quad \quad \quad \quad 
   \includegraphics[scale=0.35]{Figures/Fig6d.eps}\put(-200,150){(d)} \hfill 
  \vspace{1.7cm}  
 \includegraphics[scale=0.35]{Figures/Fig6e.eps}\put(-200,150){(e)} \quad \quad \quad \quad \quad 
   \includegraphics[scale=0.35]{Figures/Fig6f.eps}\put(-200,150){(f)} \hfill
    \caption{Shear rate scaled segregation velocity (empty symbols) and diffusional velocity (filled symbols) with the concentration of other species in a mixture for two different time durations $t_0 \leq t \leq 2\tau$ (left column) and $2\tau \leq t \leq 4\tau$ (right column). Results for the same size and same density particle mixture are shown in (a) and (b). Results for a binary mixture of density ratio $\rho=3.0$ starting from a mixed initial configuration are shown in (c) and (d). (e) and (f) show data for heavy near base initial configuration. Blue circles, and green squares represent the light and heavy species respectively.}
    \label{fig:Onlydiff-withDiff}
\end{figure*}

\begin{figure*}[!htp]
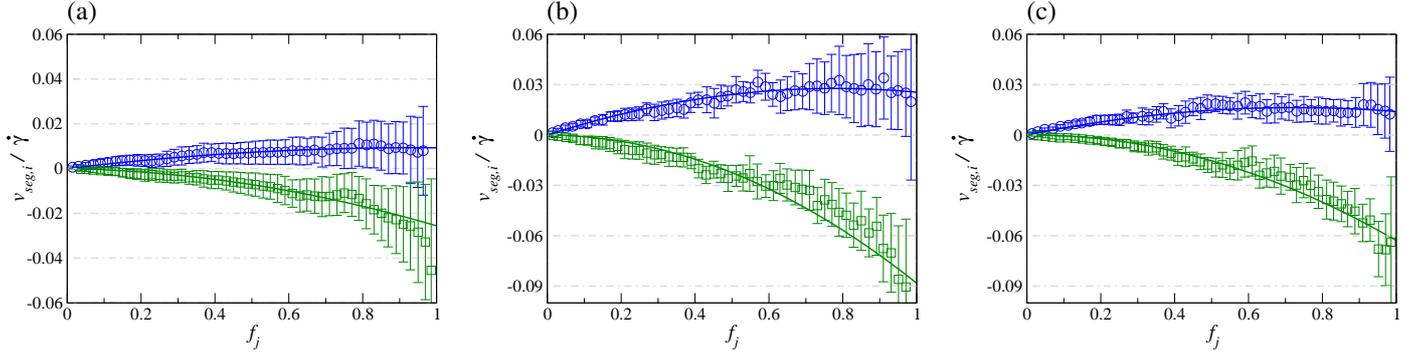

    \centering
    \includegraphics[scale=0.3]{Figures/Fig7a.eps}\put(-140,125){(a)}\hfill
    \includegraphics[scale=0.3]{Figures/Fig7b.eps}\put(-140,125){(b)}\hfill 
     \includegraphics[scale=0.3]{Figures/Fig7c.eps}\put(-140,125){(c)}\hfill
    \caption{Shear rate scaled segregation velocity variation with concentration for binary mixture having for density ratio (a) $\rho=1.5$; (b) $\rho=3.0$ at inclination angle $\theta = 25\degree$, (c)  $\rho=3.0$ at inclination angle $\theta = 29\degree$. Blue symbols represent the light species, and green symbol represents the heavy species. Blue and green solid lines represent the quadratic fit to the data $v_{seg,i}/\dot{\gamma}= d f_j (A_{ij} + B_{ij} f_j$).}
    \label{fig:SegVelocity-diffusion}
\end{figure*}

The rationale for the addition of the last term can be understood by multiplying the above equation with $f_i$. The equation then states that the flux of the species $i$ in the $y$-direction $ f_i v_{seg,i}$ is due to the combined effect of convective flux $f_i (v_i-v_{mix})$ and the diffusion flux $ D\partial f_i / \partial y $. In the case of significant size or density differences between the species, the contribution of the convective flux is much larger than the diffusive flux, and hence ignoring the latter leads to little error.
It is important to note that the above discussion pertains to the empirical measurement of the segregation velocity using the DEM simulations. The previous studies used only the relative velocity of the species with respect to the mixture velocity to measure the segregation velocity at any location. The revised method accounts for the correction due to the presence of diffusion. Given the focus of the present study, we investigate the role of convective and diffusional contributions to the segregation velocity calculations in detail. 

\textcolor{black}{To better understand this, we consider a system in which only diffusion is present, and the segregation effect is absent. We consider mono-disperse particles, identical in all aspects but tagged as two different species flowing over an inclined surface. The particles in the lower portion are tagged as species $1$ and the upper portion particles are tagged as species $2$.}
In this case, the sizes and densities of both species are equal, and hence we expect the segregation velocity to be zero and only diffusion velocity to be present. 
The figure~\ref{fig:Onlydiff-withDiff}a and \ref{fig:Onlydiff-withDiff}b show the shear rate scaled segregation velocity $\frac{v_{seg,i}}{\dot{\gamma}}$ where $v_{seg,i}$ is computed using equation~\ref{eq:duan_Vseg}, as well as shear rate scaled diffusional velocity $\frac{v_{diff,i}}{\dot{\gamma}}$ (where $ v_{diff,i} = \frac{1}{f_i} D \frac{\partial f_i}{\partial y}$). Filled symbols represent the diffusional velocity and empty symbols represent the segregation velocity. Evidently, we find that the segregation velocity using equation~\ref{eq:duan_Vseg} is close to zero at all times. This indeed appears to be the case for not only the first two time scales (figure~\ref{fig:Onlydiff-withDiff}a) but also the next two time scales (figure~\ref{fig:Onlydiff-withDiff}b). Diffusional velocity, represented with filled symbols, is noticeable only for initial times $t_0 \leq t \leq 2\tau$ (figure~\ref{fig:Onlydiff-withDiff}a). It becomes close to zero for time $2\tau \leq t \leq 4\tau$ as both identical species mix together, and the concentration gradient becomes zero (figure~\ref{fig:Onlydiff-withDiff}b). 
As in figure~\ref{fig:density-Raw-data}, the data shown in figure~\ref{fig:Onlydiff-withDiff} are obtained by considering five different mixtures of composition $20\%, 40\%, 50\%, 60\%$ and $80\%$ of one of the species. In figures \ref{fig:Onlydiff-withDiff}a and \ref{fig:Onlydiff-withDiff}b, we also use the data for $10\%$ and $90\%$ compositions to obtain the entire range of concentration from $0$ to $1$.

Figure \ref{fig:Onlydiff-withDiff}c \& \ref{fig:Onlydiff-withDiff}d show the diffusional velocity along with the segregation velocity (given by equation \ref{eq:duan_Vseg}), \textcolor{black}{both} scaled by the shear rate. The data are shown for a binary mixture of density ratio $\rho = 3.0$ flowing at $\theta = 25\degree$.  
During the $t_0 \leq t \leq 2\tau$ time interval, the diffusional velocity is found to be close to zero (figure \ref{fig:Onlydiff-withDiff}c) \textcolor{black}{since we start from a uniformly mixed configuration}. Hence the segregation velocity arises primarily due to the convective motion of the species relative to the mixture. 
Figure \ref{fig:Onlydiff-withDiff}d shows the same data for a time interval ranging from two to four time scales, i.e., $2\tau \leq t \leq 4\tau$. 
In contrast to figure \ref{fig:Onlydiff-withDiff}c, the diffusional velocities of both species appear to be almost equal to the segregation  \textcolor{black}{velocities confirming that the convective contribution $v_i-v_{mix}$ (not shown here) to the segregation velocity is negligible.} Figures \ref{fig:Onlydiff-withDiff}e \& \ref{fig:Onlydiff-withDiff}f, show the same data for a mixture starting from a completely segregated initial configuration with heavy particles near the base. As before, figure \ref{fig:Onlydiff-withDiff}e shows the data for time up to two-time scale, while figure \ref{fig:Onlydiff-withDiff}f shows the data for a time interval ranging from two to four time scales. In contrast to figure~\ref{fig:Onlydiff-withDiff}c \& \ref{fig:Onlydiff-withDiff}d for mixed configuration, it is evident that convective flux is negligible not only for time $2\tau \leq t \leq 4\tau$ but also for time $t_0 \leq t \leq 2\tau$.

The observations made in figure \ref{fig:Onlydiff-withDiff}c-\ref{fig:Onlydiff-withDiff}f can be understood as follows. Starting from a well-mixed configuration, the concentration gradient $\partial f_i / \partial y$ \textcolor{black}{and hence diffusional contribution} remains close to zero (figure~\ref{fig:Onlydiff-withDiff}c) initially, and the segregation occurs entirely due to the relative motion of the species with respect to each other. This relative motion leads to the build-up of concentration gradient  $\partial f_i / \partial y$ across the layer, which leads to increasing importance of the diffusional contribution at later times (figure~\ref{fig:Onlydiff-withDiff}d). 
\textcolor{black}{In the case of starting from an initial configuration of segregated case, the high concentration gradients exist since beginning and hence the behaviour in figure~\ref{fig:Onlydiff-withDiff}e is similar to figure~\ref{fig:Onlydiff-withDiff}d. As the final steady state configuration also has heavy particles concentrated near the base, no difference in the figure~\ref{fig:Onlydiff-withDiff}e and figure~\ref{fig:Onlydiff-withDiff}f noticeable.} 
Results shown in figure \ref{fig:Onlydiff-withDiff}c-\ref{fig:Onlydiff-withDiff}f confirm that the assumption of diffusional contribution being small in case of density segregation is not correct and the relative importance of the diffusion flux and convective flux depends upon the initial configuration and may also change as the flow evolves.

Figure \ref{fig:SegVelocity-diffusion} shows the variation of the shear rate scaled segregation velocity with the concentration of other species for the initially mixed configuration. The segregation velocity $v_{seg,i}$ is computed using equation \ref{eq:duan_Vseg} by utilizing the data for time duration $2\tau \leq t \leq 4\tau$. Green square symbols represent the DEM data for heavy species, and blue circles represent the light species. Figures \ref{fig:SegVelocity-diffusion}a - \ref{fig:SegVelocity-diffusion}b show the effect of density ratio on shear rate scaled velocity variations at inclination angle $\theta = 25\degree$.
Figure \ref{fig:SegVelocity-diffusion}a corresponds to density ratio $\rho = 1.5$ and \ref{fig:SegVelocity-diffusion}b corresponds to $\rho = 3.0$. Figure \ref{fig:SegVelocity-diffusion}c shows data for density ratio $\rho = 3.0$ at $\theta = 29\degree$.
Comparing the magnitude of $v_{seg,i}/\dot \gamma$ in figures \ref{fig:SegVelocity-diffusion}a and \ref{fig:SegVelocity-diffusion}b, it is clear that it increases with increase in the density ratio.
Comparison of figures \ref{fig:SegVelocity-diffusion}b and \ref{fig:SegVelocity-diffusion}c shows the effect of inclination angle at a given density ratio $\rho = 3.0$.
An increase in the inclination angle from $\theta = 25\degree$ to $\theta = 29\degree$ leads to a slight reduction in the shear rate scaled segregation velocity. The solid lines correspond to a quadratic fit confirming that the shear rate-scaled segregation velocity exhibits a quadratic variation with concentration. We use the segregation velocity relation proposed by  \cite{jones2018asymmetric} and \cite{Duan2021} to fit the DEM data shown in figure \ref{fig:SegVelocity-diffusion}. The authors used the following expression relating shear rate segregation velocity with concentration,
\begin{equation}
\centering
    v_{seg,i}/\dot{\gamma} = d f_j (A_{ij} + B_{ij} f_j ) ,
    \label{eq:vp-density-quadratic}
\end{equation}
where $A_{ij}$ and $B_{ij}$ are the empirical parameters. We report the values of these parameters for both species in table \ref{tab:Empriical_qua_theta_density}. We observe that the magnitudes of $A_{LH}$ and $B_{LH}$ for light species increase with increase in the density ratio, and decreases with increase in the inclination angle. This is consistent with the findings of \cite{jones2018asymmetric} and \cite{Duan2021}. 
From these results, we conclude that quadratic variation of shear rate scaled segregation velocity with concentration is indeed observed in the case of chute flow as well. However, the dependency on inclination angle $\theta$ is also observable \textcolor{black}{since the parameters $A_{ij}$ \& $B_{ij}$ for $\theta = 25\degree$ \& $\theta=29\degree$ differ substantially from each other. While increase in $\theta$ leads to higher shear rates, the model is unable to account for this dependency on $\theta$ using the shear rate scaling.}
 \begin{table}[H]
    \centering
    \begin{tabular}{@{}m{4em}m{4em}ccccc@{}}
    \toprule
    Density ratio ($\rho$) & inclination angle ($\theta$) & \textbf{$A_{LH}$} & \textbf{$B_{LH}$} & \textbf{$A_{HL}$} & \textbf{$B_{HL}$} \\
    \midrule
    \textbf{$\rho = 1.5$} &  \textbf{$\theta = 25\degree$} & 0.022 & -0.013 & 	 -0.001 & -0.026 \\
    \textbf{$\rho = 3.0$} &  \textbf{$\theta = 25\degree$} & 0.082 & -0.054 &	 -0.003 & -0.083 \\
    \textbf{$\rho = 3.0$} &  \textbf{$\theta = 29\degree$} & 0.047 & -0.028 & 	 -0.004 & -0.057 \\
    \bottomrule
    \end{tabular} 
    \caption{Empirical parameters for the species \textcolor{black}{obtained} using quadratic fitting (equation \ref{eq:vp-density-quadratic}) of $v_{seg,i}/\dot \gamma$ vs $f_j$ data for binary mixture, shown in figure \ref{fig:SegVelocity-diffusion}.}  \label{tab:Empriical_qua_theta_density}
\end{table}

\begin{figure*}[!htb]
    \centering
  \quad \quad   \includegraphics[scale=0.22]{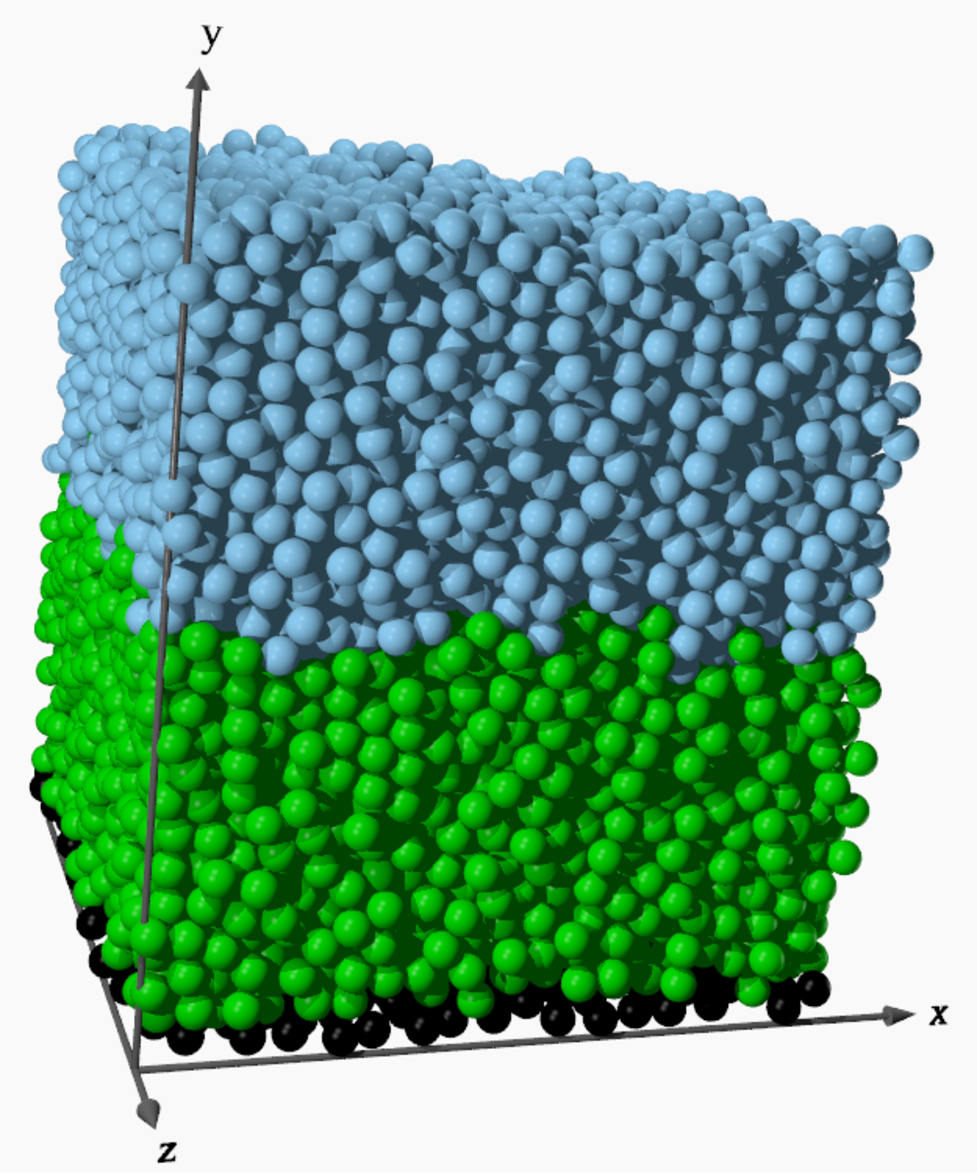}\put(-110,130){(a)}\put(-120,60){\rotatebox{90}{HNB}}\quad \quad
    \includegraphics[scale=0.30]{Figures/Fig8c.eps}\put(-140,130){(c)} \quad \quad 
   \includegraphics[scale=0.3]{Figures/Fig8e.eps}\put(-140,125){(e)}\quad \quad \hfill
   \vspace{1.7cm}
      \includegraphics[scale=0.22]{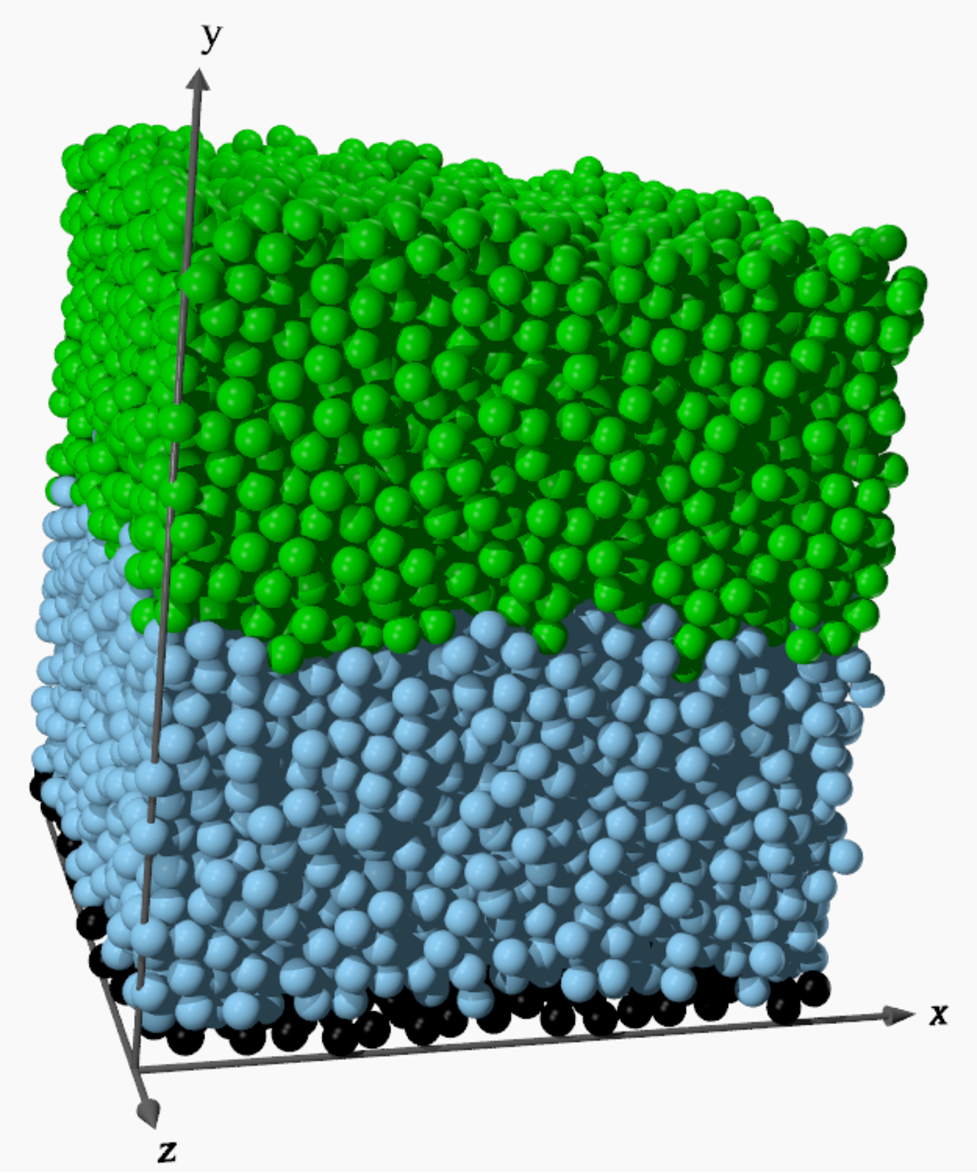}\put(-110,130){(b)}\put(-120,60){\rotatebox{90}{LNB}}\quad \quad
       \includegraphics[scale=0.30]{Figures/Fig8d.eps}\put(-140,130){(d)}\quad \quad 
        \includegraphics[scale=0.3]{Figures/Fig8f.eps}\put(-140,125){(f)}
    \caption{Snapshots showing an initial configuration of (a) Heavy near base (HNB) case and (b) Light near base (LNB) case for a binary mixture of equal compositions of each species at an inclination angle $\theta = 25^\circ$  with density ratio $\rho_{H}:\rho_{L} = 3:1$. Black spheres represent the bumpy base.  Shear rate scaled segregation velocity (neglecting diffusional velocity) \textcolor{black}{used in linear empirical model} for the (c) HNB case and (d) LNB case (error bars shown only for alternate data points for the sake of clarity). 
   The data are taken for $t_0 \leq t_f \leq 2.0\tau$. Shear rate scaled segregation velocity (accounting for diffusional velocity) \textcolor{black}{used in quadratic empirical model} for the (e) HNB case and (f) LNB case. The data are taken for $2.0 \tau \leq t_f \leq 4.0\tau$.}
    \label{fig:den-conf}
\end{figure*} 
We next explore the effect of initial configuration on the empirical parameters estimated using the two \textcolor{black}{segregation models}.
We estimate the empirical parameters for binary granular mixtures starting from two different initial configurations.
We follow the terminology used by \cite{sahu_kumawat_agrawal_tripathi_2023}.
The configuration with the heavy species initially concentrated near the base is referred to as the heavy near base (HNB) case. A snapshot for the HNB case is shown in figure \ref{fig:den-conf}a. 
The light near base (LNB) case, depicted in figure \ref{fig:den-conf}b, consists of the low-density species concentrated entirely near the base and the high-density species entirely near the free surface. In figures \ref{fig:den-conf}a and \ref{fig:den-conf}b, the green spheres represent the high-density particles, and blue spheres represent the low-density particles. The inclined surface has an angle of $25\degree$ and the mixture has a density ratio $\rho = 3.0$.
Figures \ref{fig:den-conf}c and \ref{fig:den-conf}d show the variation of $v_{seg,i}/\dot \gamma$ with the concentration of other species $f_j$ for HNB and LNB cases, respectively \textcolor{black}{for the linear empirical model.} $v_{seg,i}$ (computed using equation \ref{eq:segVel_vp-v}) divided by the shear rate is plotted with $f_j$ by utilizing the data for observation time up to twice the segregation time scale ($2 \tau$).
The magnitude of $v_{seg,i}/\dot \gamma$ for both HNB and LNB case are different from that for the mixed state (shown in figure \ref{fig:density-Raw-data}b). 
Since the HNB configuration is already close to the final segregated state, very low values of the percolation velocities are obtained. 
The shear rate scaled segregation velocity for the LNB case is approximately $10$ times higher than the HNB case. 
Using linear fit to get the empirical parameters for linear model appears to differ by nearly an order of magnitude for the two configurations. Further, the data for the LNB case does not seem to be described well using the linear fit for $f_j \geq 0.50$.
These results confirm the dependency of initial configuration on the parameter estimation using the linear empirical model. 

Figure~\ref{fig:den-conf}e shows the $v_{seg,i}/\dot \gamma$ calculated using equation \ref{eq:duan_Vseg} and relate it with $f_j$ using the quadratic empirical approach for HNB case. Figure \ref{fig:den-conf}f shows the same data for the LNB case. In both the figures, we employ data beyond twice the segregation time scale $2\tau$ (i.e., $2\tau \leq t \leq 4\tau$). The magnitude of the $v_{seg,i}/\dot \gamma$ in both the configurations is found to be comparable to each other as well as to the mixed case discussed in figure \ref{fig:SegVelocity-diffusion}b. Fitting equation \ref{eq:vp-density-quadratic} to the data reveals that the values of the empirical parameters $A_{ij}$ and $B_{ij}$ for both species are \textcolor{black}{slightly different from each other. However, the predicted segregation velocity seems to differ only for large values of $f_j$.} Thus we can conclude that the initial configuration doesn't play as significant role in empirical parameter estimation in the case of the quadratic empirical model \textcolor{black}{as in the linear case.} 

\section{Comparison of empirical and particle force-based segregation models}
\label{sec:Comp_Emp_and_Ptcle}
The studies for heap flow and rotating cylinders suggest that the empirical models of segregation velocity coupled with the convection-diffusion-segregation equation can predict the concentration profile in good agreement with DEM simulations and experiments. In this section, we explore the applicability of the empirical approach for periodic chute flows.
We solve the steady state $1D$ diffusion-segregation equation \citep{tripathi2013density,sahu_kumawat_agrawal_tripathi_2023} along with the momentum balance equations and the inertial number-based rheology following the approach \cite{tripathi2011rheology} and \cite{sahu_kumawat_agrawal_tripathi_2023}. Both empirical as well as particle force-based segregation models have been utilized to predict the steady-state concentration profiles theoretically. 
\subsection{Theory}
\label{subsec:theory}
We consider a fully developed flow over an inclined surface which is simulated by employing a periodic chute in the DEM simulations. For this system, the concentration gradient along the flow direction vanishes, and hence, the convection term in the convection-diffusion-segregation equation (\cite{xiao2016modelling,Duan2021}) becomes zero. At steady state, the transient term also vanishes, and the equation reduces to a simple flux balance equation for the species (\cite{tripathi2013density}, \cite{sahu_kumawat_agrawal_tripathi_2023}),
\begin{equation}
    \Phi_{seg,i} + \Phi_{diff,i} = 0.
    \label{eq:steadyflux}
\end{equation}
Using the expression for the diffusion flux $\Phi_{diff,i} = -D\frac{\partial f_i}{\partial y}$ and segregation flux $\Phi_{seg,i} = v_{seg,i}f_i $, we obtain an ordinary differential equation
\begin{equation}
   v_{seg,i}f_i -  D\frac{df_i}{d y} =  0,
   \label{eq:flux-balance}
\end{equation}
where $v_{seg,i}$ is the segregation velocity and 
$D$ is the diffusivity. Several studies, including \cite{fan2014modelling}, \cite{schlick2016}, \cite{deng2018continuum}, and \cite{Deng2020} utilized a constant value of the diffusivity in the theoretical predictions. 
Utilizing a constant diffusivity in the empirical approach to solve equation \ref{eq:flux-balance} would require the knowledge of $\dot \gamma$ due to the dependence of percolation velocity $v_{seg,i}$ on the local shear rate $\dot \gamma$. For this, one needs to incorporate information about flow kinematics for predicting segregation. DEM simulations confirm that the average value of diffusivity depends on inclination angle $\theta$ (\textcolor{black}{see} Figure 8 by \citet{tripathi2013density}) and hence separate diffusivity values need to be used for each inclination angle to get reasonable predictions. \textcolor{black}{ Table~\ref{table:Diffusivity-value} in the appendix shows that a very weak dependency of the average diffusivity on the mixture composition is observed. However, the average diffusivity values for $\theta=25^\circ$ and $\theta=29^\circ$ differ significantly from each other. Figure~\ref{fig:diff_supfig} confirms that the predictions using the particle force-based theory also differ significantly if the diffusivity value corresponding to the particular inclination angle is not used.}

Variation of the diffusivity with the shear rate has been reported by many researchers (\cite{bridgwater1985particle, utter2004self, tripathi2013density, Fan2015, fry2019, sahu_kumawat_agrawal_tripathi_2023}). It is observed that the diffusivity ($D$) varies linearly with the shear rate, $D = b \dot \gamma d^2_{mix}$, where $d_{mix}$ is the \textcolor{black}{local} volume average diameter. $b$ is a constant parameter obtained using DEM simulation. We use $b = 0.041$, obtained by \cite{tripathi2013density} and \cite{ sahu_kumawat_agrawal_tripathi_2023} in the case of density segregation. For different density and same size particle mixtures, $d_{mix} = d$. By utilizing the variation of diffusivity on the local shear rate and concentration using relation $D = b \dot \gamma d^2_{mix}$, the need for the detailed information about flow kinematics is eliminated, and the equation can be solved analytically. The usage of diffusivity linearly scaling with the shear rate automatically takes care of diffusivity variation with inclination and hence should be preferred.
In this study, we account for the shear rate dependence of the diffusivity and use this relation to obtain analytical solutions of equation~\ref{eq:flux-balance} for both linear and quadratic empirical relation of $v_{seg,i}$. For the linear segregation model (given by equation~\ref{eq:vp_density}) equation~\ref{eq:flux-balance} simplifies to
\begin{equation}
   b \dot{\gamma} d^2 \frac{d f_i}{d y} =  S_D(i,j)\dot{\gamma} f_j f_i. 
   \label{eq:flux-with-linearEmp}
\end{equation}
Using $f_j = 1 - f_i$ for the binary mixture case and integrating equation~\ref{eq:flux-with-linearEmp} with boundary condition $f_i( y = 0) = f_{i0}$, we get
\begin{equation}
  f_i = \frac{f_{i0} \exp{\left( \frac{S_D}{b d^2} y \right)} }{1 - f_{i0} \left(1 - \exp\left(\frac{S_D}{b d^2} y\right)\right)}.
   \label{eq:analytical-linear}
\end{equation}
Equation \ref{eq:analytical-linear} describes the variation of species concentration $f_i$ with $y$ across the layer using an explicit function for the linear empirical segregation model.
On the other hand, using quadratic empirical relation (given by equation~\ref{eq:vp-density-quadratic}) in the flux balance equation (equation \ref{eq:flux-balance}), we get 
\begin{equation}
   b \dot{\gamma} d^2 \frac{d f_i}{d y} =  \dot{\gamma} d f_j (A_{ij} + B_{ij} f_j ) f_i, 
   \label{eq:flux-with-QuadraticEmp}
\end{equation}
where $f_j = 1 - f_i$ is the concentration of other species. Integrating equation~\ref{eq:flux-with-QuadraticEmp} with boundary condition $f_i( y = 0) = f_{i0}$, we get an implicit equation for $f_i$ given as
\begin{dmath}
    y  = \frac{1}{b d}\left[ \frac{B_{ij}}{A_{ij}(A_{ij} + B_{ij})} \log \left (\frac{A_{ij} + B_{ij}  - B_{ij} f_{ij}}{A_{ij} + B_{ij} - B_{ij} f_{i0}}\right) \quad \quad \quad \quad \quad \quad + \frac{1}{A_{ij} + B_{ij}} \log \left(\frac{f_i}{f_{i0}}\right) - \frac{1}{A_{ij}} \log \left(\frac{1 - f_i}{1 - f_{i0}}\right) \right].
    \label{eq:analytical-quad}
\end{dmath}
Both equation~\ref{eq:analytical-linear} and equation~\ref{eq:analytical-quad} require the knowledge of $f_{i0}$ i.e., the value of $f_i$ at $y=0$ to predict the concentration variation across the layer. This value, however, is not known apriori. In order to obtain the value of $f_{i0}$, we make use of the overall species balance across the flowing layer.
\textcolor{black}{For this mass balance,} the depth-averaged concentration of the species in the layer must be equal to the total mixture composition $f^T_i$. In other words, the integral of the species concentration $f_i(y)$ across the flowing layer divided by the layer thickness $H$ must be equal to the total composition of the species $i$, i.e., 
\begin{equation}
    \frac{1}{H} \int_0^{H} f_i dy - f_i^T = 0.
     \label{eq:initial-condn}
\end{equation}
Equation~\ref{eq:initial-condn} can be solved analytically for the linear empirical model. Specifically, we use equation~\ref{eq:analytical-linear} in the equation~\ref{eq:initial-condn} to get 
\begin{equation}
 \frac{1}{H} \int_0^{H} \frac{f_{i0}\exp( \frac{S_D}{b d^2} y )}{1-f_{i0} + f_{i0}\exp( \frac{S_D}{b d^2}y)} dy =  f_i^T.
  \label{eq:Int-bc-linear-emp}
\end{equation}
Equation~\ref{eq:Int-bc-linear-emp} can be easily integrated to get the value of the unknown concentration $f_i$ at $y = 0$, i.e., $f_{i0}$ 
\begin{equation}
  f_{i0} = \frac{1-\exp{(\frac{S_D}{b d^2}H f_i^T)} }{1 - \exp(\frac{S_D}{b d^2}H)}. 
\end{equation}
In the case of the quadratic model analytical solution \textcolor{black}{for $f_{i0}$} can't be obtained. In this case, we obtain $f_{i0}$ iteratively by following the approach used by the \cite{tripathi2013density}. We start by assuming an initial guess value of $f_{i0}$ at $y = 0$ and solve the equation~\ref{eq:analytical-quad} to compute the species concentration $f_i$ across the layer height $H$. Subsequently, we numerically \textcolor{black}{calculate} the $LHS$ of equation~\ref{eq:initial-condn}, which should be equal to $0$ to ensure the species mass balance. Numerically this condition is assumed to be satisfied if the absolute value of the $LHS$ of equation \ref{eq:initial-condn} turns out to be less than the tolerance value. In this work, we use a tolerance value of $10^{-4}$. Given that $f^T_i \geq 0.1$, this tolerance value leads to less than $0.1\%$ error in the mass balance. If this condition is not satisfied, we update the value of $f_{i0}$ by a small increment \textcolor{black}{(or decrement) of} $\Delta f_0 = 10^{-5}$. We repeat this process until the tolerance criteria is satisfied to obtain $f_{i0}$.

\textcolor{black}{While the flux balance equations for the binary mixture case (equations~\ref{eq:flux-with-linearEmp} and \ref{eq:analytical-quad}) can be solved analytically, the equations for mixtures with more than two components require a numerical solution.}
We note that the linear empirical model has been extended for multi-component mixtures differing in density by \cite{deng2018continuum}. The particle force-based segregation model has also been extended for multi-component mixtures by \cite{sahu_kumawat_agrawal_tripathi_2023}. However, the quadratic empirical model has not been extended for multi-component density segregation. Studies involving multi-component mixtures differing in size (\cite{gray_ancey_2011,deng2019modeling}) as well density (\cite{deng2018continuum,sahu_kumawat_agrawal_tripathi_2023}) suggest that the segregation flux $\phi_{seg,i}^{Multi}$ of a species $i$ in a multi-component mixture of $N$ species can be written as the summation of the binary segregation fluxes with respect to all the other species in the mixture, i.e., $\phi_{seg,i}^{Multi} = \sum_{j=1,j \neq i }^{N} \phi_{seg,i}^{Binary} (i,j)$. The species flux balance (equation~\ref{eq:steadyflux}) for multi-component mixtures can be written as 
\begin{equation}
    - D \frac{\partial f_i}{\partial y} + \sum_{j=1,j \neq i }^{N} \phi_{seg,i}^{Binary} (i,j) = 0.
    \label{eq:multi-diff-balance}
\end{equation}
Employing the linear empirical model given by equation~\ref{eq:vp_density}, i.e., $v_{seg,i} = S_D(i,j) \dot \gamma f_j$ and $\phi_{seg,i} = v_{seg,i} f_i$, we have $\phi_{seg,i}^{Binary} (i,j) = S_D(i,j) \dot \gamma f_j f_i$. Since $D = b \dot \gamma d_{mix}^2$, the dependency on the shear rate is eliminated \textcolor{black}{for the empirical model} and equation~\ref{eq:multi-diff-balance}  simplifies to 
\begin{equation}
    \dfrac{\partial f_i}{\partial y} = \frac{1}{b d^2}\sum_{j=1,j \neq i }^{N} S_D(i,j) f_i f_j.
    \label{eq:multi-comp-linear-diff}
\end{equation}
In the case of the quadratic empirical model $\phi_{seg,i}^{Binary} = (A_{ij} + B_{ij} f_j) f_j f_i$ and the equation~\ref{eq:multi-diff-balance} simplifies to 
\begin{equation}
    \dfrac{\partial f_i}{\partial y} = \frac{1}{b d}\sum_{j=1,j \neq i }^{N} (A_{ij} + B_{ij} f_j) f_i f_j.
    \label{eq:multi-comp-quadratic-diff}
\end{equation}

\begin{algorithm*}[!htb]
\SetAlgoLined
 \For{$i\gets1$ \KwTo $N-1$}{
 Initialize $f_i(y) = f_i^T$ at all $y$\;
 Initialize $f_{i0}$\;
 }
define  $error = 1$; $tolerance = 10^{-4}$; $\Delta f_0 = 10^{-5}$\;

 \While{$error > tolerance$ }{
 for use in ODE45 solver\; 
 
 \For{$i\gets1$ \KwTo $N-1$}{
 Define $f_i(y)$ in symbolic form\;
}
 \For{$i\gets1$ \KwTo $N-1$}{
  Solve equations~\ref{eq:multi-comp-linear-diff} (\ref{eq:multi-comp-quadratic-diff}) using ode45 with boundary condition $f_i(y=0) = f_{i0}$ to compute new $f_i(y)$\;
 
  Compute difference between the calculated and actual total composition of the species using \ref{eq:initial-condn}; 
    }
    Compute $f_N=1-\sum\limits_{i=1}^{N-1} f_i$ \;
    Define $error = \sqrt{\sum\limits_{i=1}^{N-1}\left|\dfrac{1}{H} \int_{0}^{H} f_{i} d y - f_i^T\right|^2}$ \;

 \For{$i\gets1$ \KwTo $N-1$}{    
  \eIf{$\dfrac{1}{H} \int_{0}^{H} f_{i} d y < f_i^T$}{
   $f_{i0} = f_{i0} + \Delta f_0 $\;
   }{
    $f_{i0} = f_{i0} - \Delta f_0 $\;
  }
  }
 }
 \caption{Algorithm used for predicting concentration of $N$ component mixture using empirical segregation model}
 \label{algo1}
\end{algorithm*}

Equations~\ref{eq:multi-comp-linear-diff} and \ref{eq:multi-comp-quadratic-diff} represent the spatial variation of species $i$ in a mixture of $N$ components. Hence $i$ can take any value from $1$ to $N$. Since $\sum^N_{i=1} f_i = 1$, we get
$N - 1$ inter-coupled, non-linear differential equations that need to be solved simultaneously.
We solve equations~\ref{eq:multi-comp-linear-diff} and \ref{eq:multi-comp-quadratic-diff} numerically using $ode45$ in MATLAB with an initial guess of $f_{i0}$ by following the Algorithm \ref{algo1}. \textcolor{black}{The iterative method is employed in the quadratic approach to obtain $f_{i0}$.} The empirical parameters are estimated for different combinations of density ratios.
\textcolor{black}{In the linear empirical approach, the relationship of $S_D$ with density ratios $S_D(i,j) = C_D d \ln{\rho_j/\rho_i}$ facilitates solution for any generic pair of $\rho_i$ and $\rho_j$. However, for the quadratic model, parameters $A_{ij}$ and $B_{ij}$ are obtained for a particular density ratio.} 

Note that it is possible to obtain the relation of quadratic empirical model parameters $A_{ij}$ and $B_{ij}$ with density ratio as reported by \cite{Duan2021}. However, given the focus of the paper, we use the $A_{ij}$ and $B_{ij}$ for the specific combinations of density ratios corresponding to the binary mixture of the same density ratio. Using the equations described above, we obtain the concentration profiles of species across the layer for empirical segregation models (linear as well quadratic). The predictions for the particle force-based theory for binary mixtures are done following the approach of  \cite{tripathi2013density}. Specifically, we use the algorithm given in Table 5 of their work.
\textcolor{black}{For ternary mixtures, we follow algorithm given by \cite{sahu_kumawat_agrawal_tripathi_2023}.}
We utilize the same set of parameters as used in these references and the interested reader is encouraged to look at these references for more details. 

\subsection{Comparison of theoretical predictions}
Next, we compare the theoretical prediction using the linear empirical model of \cite{xiao2016modelling} and quadratic empirical model of \cite{Duan2021} with the particle force-based theory proposed by  \cite{tripathi2013density} and \cite{sahu_kumawat_agrawal_tripathi_2023}. 
As shown in section \S \ref{sec:empirical-parameter}, the empirical parameters for the mixed configuration obtained using the linear segregation model are more reliable in comparison to LNB and HNB configurations. Hence, we utilize the empirical parameters obtained for the mixed configuration in these theoretical predictions. 
\textcolor{black}{While the empirical parameters obtained using the quadratic approach are found to be relatively less dependent on the initial configuration, we use the parameters for mixed configuration for quadratic model as well.} 

\begin{figure*}[!htb]
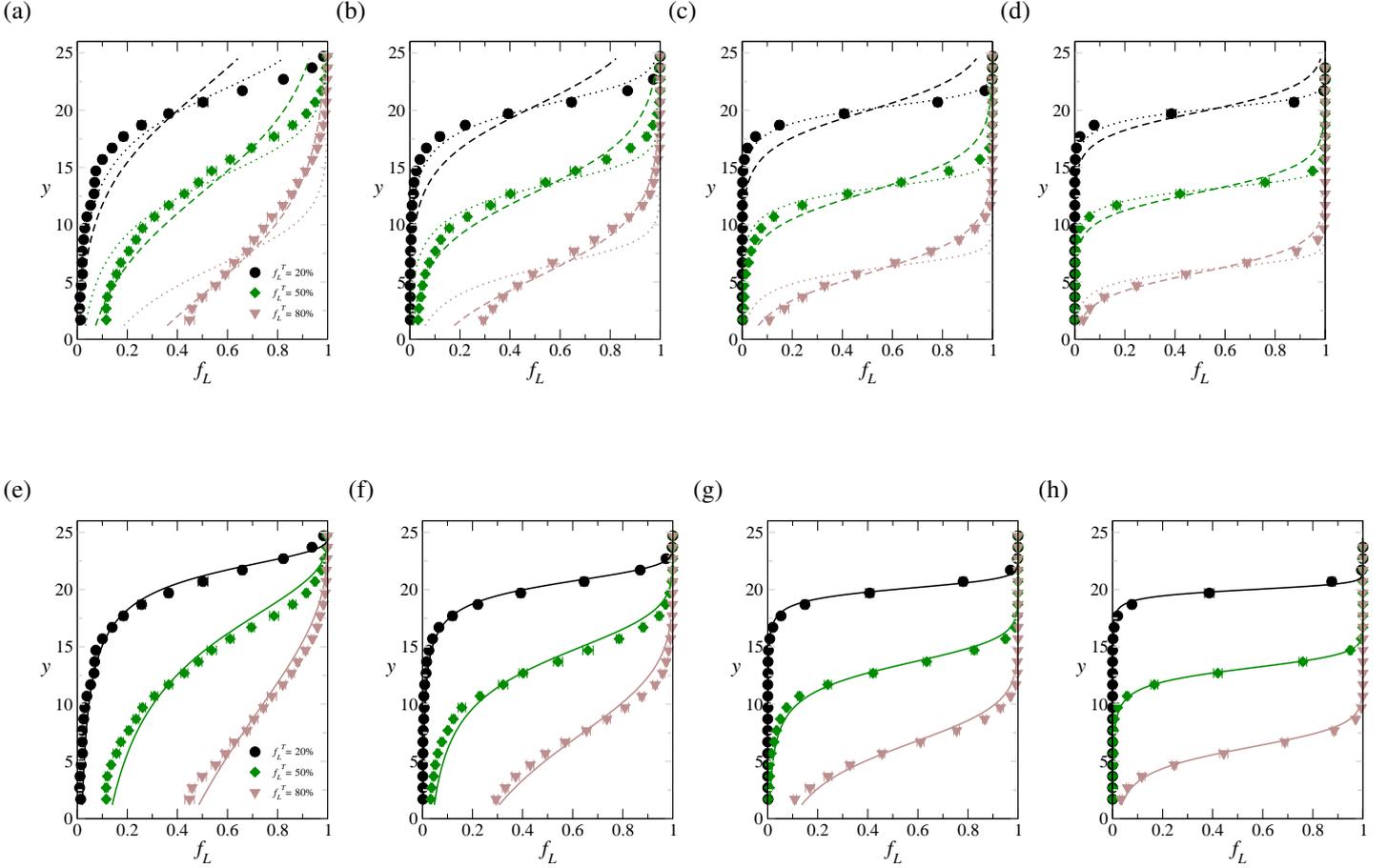

    \includegraphics[scale=0.32]{Figures/Fig9a.eps}\put(-130,145){(a)}\hfill 
     \includegraphics[scale=0.32]{Figures/Fig9b.eps}\put(-130,145){(b)}\hfill 
      \includegraphics[scale=0.32]{Figures/Fig9c.eps}\put(-130,145){(c)}\hfill 
       \includegraphics[scale=0.32]{Figures/Fig9d.eps}\put(-130,145){(d)}\hfill  
       \vspace{1.3cm}

    \includegraphics[scale=0.32]{Figures/Fig9e.eps}\put(-130,145){(e)}\hfill 
    \includegraphics[scale=0.32]{Figures/Fig9f.eps}\put(-130,145){(f)}\hfill 
    \includegraphics[scale=0.32]{Figures/Fig9g.eps}\put(-130,145){(g)}\hfill 
    \includegraphics[scale=0.32]{Figures/Fig9h.eps}\put(-130,145){(h)}\hfill 
    \caption{(a-d) Comparison of theoretical predictions using linear (dashed lines) and quadratic (dotted) empirical approach along with (e-h) theoretical predictions of particle force-based segregation model (solid lines) with DEM simulation data (symbols). The results are shown for the three different mixture compositions $f^T_L = 0.20$ (black), $0.50$ (green) and $0.80$ (brown) for binary mixtures having density ratio $\rho=1.5$ in (a,e), $\rho=2.0$ in (b,f), $\rho = 3.0$ in (c,g), and $\rho = 5.0$ in (d,h) at inclination angle $\theta=25^\circ$. }
    \label{fig:binary-linear-quad}
\end{figure*}

Figures \ref{fig:binary-linear-quad}a - \ref{fig:binary-linear-quad}d show the comparison of the theoretical predictions using linear (dashed lines) and quadratic (dotted lines) empirical models. The steady-state DEM simulation data are shown as symbols for comparison. The concentration profiles of light species are shown for the three different mixture compositions of light species $f^T_L = 0.20$ (black), $0.50$ (green) and $0.80$ (brown).
Figure \ref{fig:binary-linear-quad}a, \ref{fig:binary-linear-quad}b, \ref{fig:binary-linear-quad}c, \& \ref{fig:binary-linear-quad}d show the results for density ratio $\rho = 1.5$, $2.0$, $3.0$ and $5.0$ respectively using the empirical model. Figure~\ref{fig:binary-linear-quad}e, \ref{fig:binary-linear-quad}f, \ref{fig:binary-linear-quad}g, and \ref{fig:binary-linear-quad}h show the predictions using the particle force-based theory using solid lines for the same density ratios. Thus each column in figure \ref{fig:binary-linear-quad} corresponds to a particular density ratio. 

Let us first consider mixtures with the composition of light species $f^T_L = 0.20$.
Figure \ref{fig:binary-linear-quad}a shows that the predictions using the linear segregation model (black dashed line) show significant differences with DEM data (black circles) at density ratios $\rho = 1.5$. Similar behaviour is observed for $\rho = 2.0$ (figure~\ref{fig:binary-linear-quad}b), $\rho = 3.0$ (figure~\ref{fig:binary-linear-quad}c), and $\rho = 5.0$ (figure~\ref{fig:binary-linear-quad}d). However, these differences in the predictions appear to be decreasing with the increase in the density ratio. 
On the other hand, the quadratic empirical model (black dotted line) \textcolor{black}{shows relatively small differences and} seems to be describing the simulation data comparatively better for the density ratio of $\rho = 1.5$ for $f^T_L = 0.20$ (figure \ref{fig:binary-linear-quad}a). At higher density ratios of $\rho = 2.0$, $\rho = 3.0$, and $\rho = 5.0$, the differences \textcolor{black}{between data \& theory get substantially reduced}, and the black dotted line follows the black circles very well, indicating that the quadratic empirical segregation model predicts segregation very well for a dilute mixture of light particles for $\rho \geq 2.0$. 
For $f^T_L = 0.50$, predictions using the quadratic approach match better with the DEM results \textcolor{black}{in comparison to linear empirical model} for higher density ratios. 
At lower density ratios, the two models appear to be similar in their agreement with the DEM simulation data.
For the mixtures having high composition of light species $f^T_L = 0.80$, the predictions using the linear empirical model (dashed lines) match with DEM data very well. However, the predictions using the quadratic empirical model show noticeable differences at low density ratios. The deviation between the data (symbols) and dotted lines decreases with increase in $\rho$.
Theoretical predictions using the particle force-based theory, shown in figures~\ref{fig:binary-linear-quad}e, \ref{fig:binary-linear-quad}f, \ref{fig:binary-linear-quad}g, and \ref{fig:binary-linear-quad}h, on the other hand are found to match well with DEM data across the entire composition range and various density ratios.
\begin{figure*}[!htb]
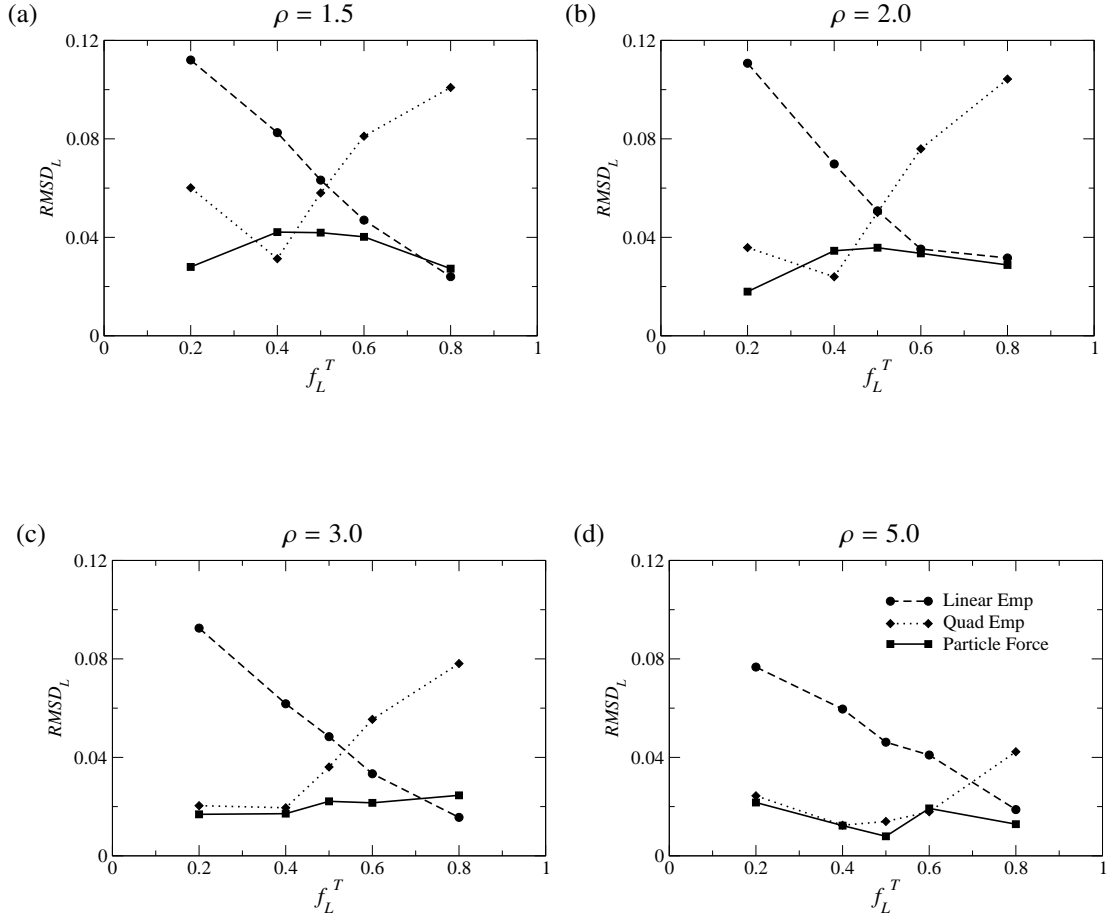

    \centering
  \hspace{2cm}  \includegraphics[scale=0.33]{Figures/Fig10a.eps}\put(-100,140){\textbf{$\rho=1.5$}}\put(-200,140){(a)} \quad \quad
    \includegraphics[scale=0.33]{Figures/Fig10b.eps}\put(-100,140){\textbf{$\rho=2.0$}}\put(-200,140){(b)} \hfill
    \vspace{1.7cm}
    \includegraphics[scale=0.33]{Figures/Fig10c.eps}\put(-100,140){\textbf{$\rho=3.0$}}\put(-200,140){(c)} \quad \quad
    \includegraphics[scale=0.33]{Figures/Fig10d.eps}\put(-100,140){\textbf{$\rho=5.0$}}\put(-200,140){(d)} 
    \caption{Variation of \textcolor{black}{light species} root mean square deviation (RMSD) value with mixture composition for three different segregation model density ratios (a) $\rho=1.5$, (b) $\rho=2.0$, (c) $\rho = 3.0$, and (d) $\rho = 5.0$. Solid lines correspond to the particle force-based model, whereas the dashed and the dotted lines correspond to the linear and quadratic empirical models, respectively.}
    \vspace{1cm}
    \label{fig:RMSD}
\end{figure*}
In order to quantify the deviations of the theoretical predictions with the simulation data, we compute the root mean square deviation (RMSD) and compare these three segregation models, namely the particle force-base model, and the linear and quadratic empirical segregation model. The RMSD value, calculated as 
\begin{equation}
    RMSD_j = \sqrt{\frac{\sum_{i=1}^{N_{bin}} \left[f^{th}_j\left(y_i\right) - f^{DEM}_j\left(y_i\right)\right]^2}{N_{bin}}},
\end{equation}
quantifies the deviation of the theoretical predictions of the concentration profile $f^{th}_j(y)$ obtained using the segregation model from that of DEM simulations data $f^{DEM}_j(y)$ for the $j^{th}$ species in the mixture. Here, $N_{bin}$ denotes the number of DEM data points in the concentration profile obtained from DEM simulations. Since we use bin sizes of $1d$ thickness to obtain DEM simulations, $N_{bin}$ equals $H/d = 24$.
Figure~\ref{fig:RMSD} shows the RMSD values for different composition mixtures at density ratio $\rho = 1.5$, $\rho = 2.0$, $\rho = 3.0$, and $\rho = 5.0$ shown in figures~\ref{fig:RMSD}a, \ref{fig:RMSD}b, \ref{fig:RMSD}c, and \ref{fig:RMSD}d respectively. As before, the dashed and dotted lines correspond to the linear and the quadratic empirical models, respectively. The solid lines correspond to the particle force-based theory. Figures~\ref{fig:RMSD}a - \ref{fig:RMSD}d show that for mixtures having $f^T_L \leq 0.50$, the RMSD values for the linear model (dashed lines) are larger than that of the quadratic model (dotted lines). However, for compositions in the range $f^T_L  \geq 0.50 $, the linear model (dashed lines) exhibits lower RMSD values compared to the quadratic model (dotted lines) except for $\rho = 5.0$. In that case, the RMSD value of the linear model is smaller than the quadratic only for $f^T_L = 0.80$.
The RMSD value decreases with increase in $f^T_L$ (total composition of light species) for the linear empirical model for all four density ratios. The RMSD for the quadratic model, on the other hand, appears to increase with increasing $f^T_L$. This quantitative measurement confirms our qualitative observations discussed above, where we observe that the linear empirical model works well for mixtures with high composition of light particles, whereas the quadratic model works better for mixtures with low composition of light species. The particle force-based theory, however, \textcolor{black}{shows that RMSD values} remain nearly constant and do not vary with the composition of the mixture. 
The RMSD value using the particle force-based model is significantly smaller compared to the other two empirical approaches across the wide range of compositions and density ratios considered in this study. 
The above results for binary mixtures confirm that particle force-based theory is able to capture the segregation much more accurately compared to the empirical model for a wide range of composition and density ratios. 

\begin{figure*}[!htb]
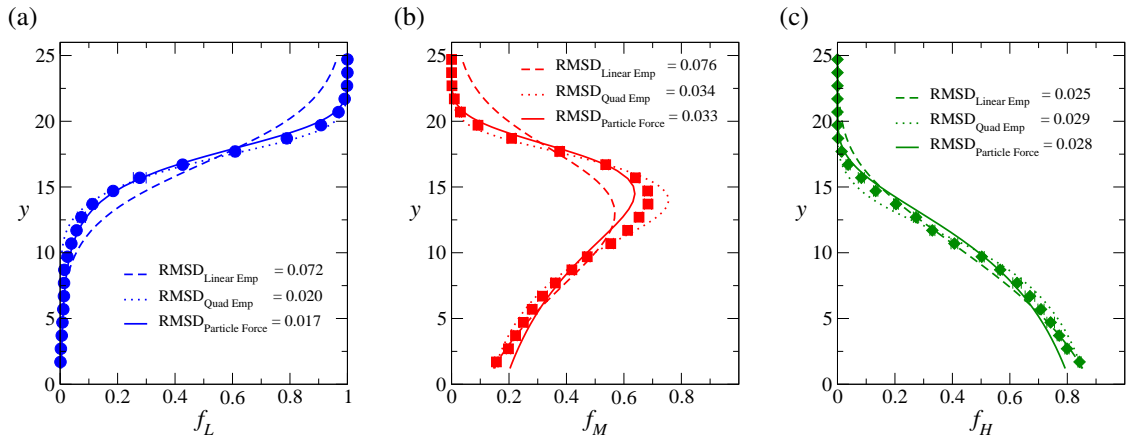

\centering
     \includegraphics[scale=0.35]{Figures/Fig11a.eps}\put(-130,155){(a)} \quad \quad
      \includegraphics[scale=0.35]{Figures/Fig11b.eps}\put(-130,155){(b)} \quad \quad
       \includegraphics[scale=0.35]{Figures/Fig11c.eps}\put(-130,155){(c)} \quad \quad
       \caption{Comparison of theoretical predictions using linear (dashed lines) and quadratic (dotted) empirical model along with the particle force-based model (solid lines) with DEM simulation data (symbols) for equal composition ternary mixture of density ratio $\rho_H : \rho_M : \rho_L = 3:2:1$. Concentration profile of (a) light-density species, (b) medium-density species, and (c) high-density species for $\theta = 25\degree$.}
        \label{fig:terany-equalComp}
\end{figure*}
\textcolor{black}{We next report results for ternary mixtures. Figure~\ref{fig:terany-equalComp} shows the steady state concentration profiles for a ternary mixture having an equal composition of the three species. The data corresponds to a density ratio of $\rho_H : \rho_M : \rho_L = 3:2:1$ and inclination angle of $\theta = 25\degree$.} Figure~\ref{fig:terany-equalComp}a shows the concentration profile of light species, whereas figures~\ref{fig:terany-equalComp}b and \ref{fig:terany-equalComp}c show that of the medium and the heavy species, respectively. Theoretical predictions using the linear empirical segregation model (dashed lines) differ from the DEM data (symbols) for light and medium species in the ternary mixture. Such differences are not observable for the heavy species. On the other hand, predictions using the particle force-based segregation model (solid lines) and quadratic empirical segregation model (dotted lines) match with the DEM results very well for all three species of the mixture. These observations can be confirmed by the RMSD values reported in the legends of figure~\ref{fig:terany-equalComp}. While the RMSD values for the heavy species are found to be nearly identical for all three segregation models, the linear empirical model RMSD values are found to be significantly higher than the other two models for both light as well as medium-density species. We also find that RMSD values for the quadratic empirical model are not very different from the RMSD values of the particle force-based theory for all three species. \begin{figure*}[!htb]
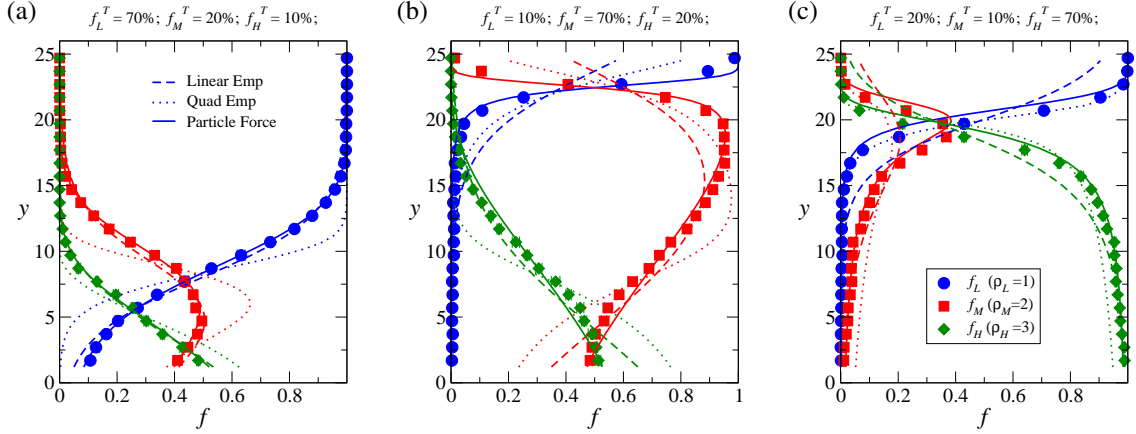

\centering
     \includegraphics[scale=0.35]{Figures/Fig12a.eps}\put(-130,155){(a)}\quad \quad
    \includegraphics[scale=0.35]{Figures/Fig12b.eps}\put(-130,155){(b)}\quad \quad
 \includegraphics[scale=0.35]{Figures/Fig12c.eps}\put(-130,155){(c)}\quad \quad
       \caption{Comparison of theoretical predictions using linear (dashed lines) and quadratic (dotted lines) empirical model along with the particle force-based model (solid lines) with DEM simulation data (symbols) for three different compositions in a ternary mixture of density ratio $\rho_H : \rho_M : \rho_L = 3:2:1$. Blue circles correspond to light species while red squares and green diamonds correspond to medium and heavy species. Concentration profiles for (a) $f^T_L = 70\%, f^T_M = 20\%, f^T_H = 10\%$, (b) $f^T_L = 10\%, f^T_M = 70\%, f^T_H = 20\%$, and (c) $f^T_L = 20\%, f^T_M = 10\%, f^T_H = 70\%$.} 
       \label{fig:ternary-conc-diffComp}
\end{figure*}
\begin{figure*}[!htb]
    \centering
      \includegraphics[scale=0.375]{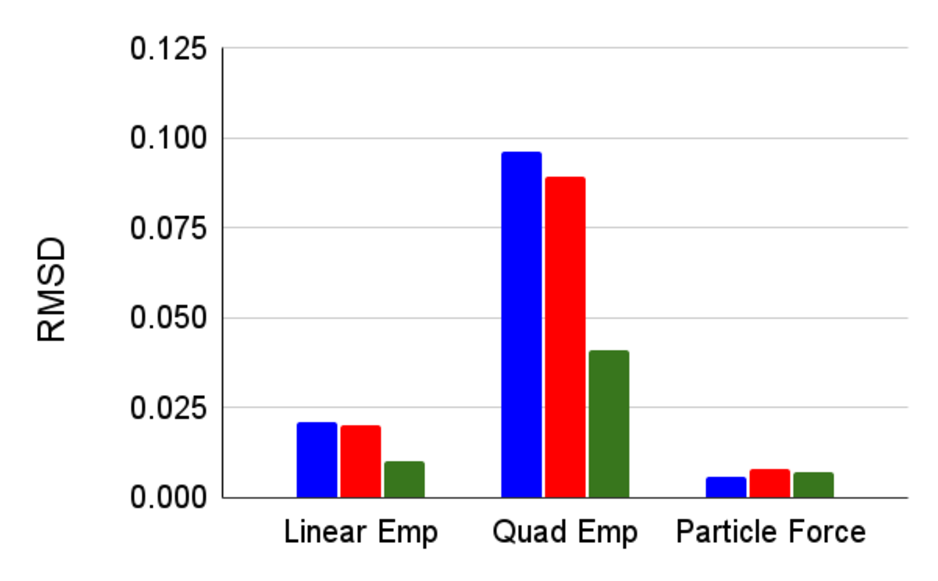}\put(-130,105){(a)}
      \includegraphics[scale=0.375]{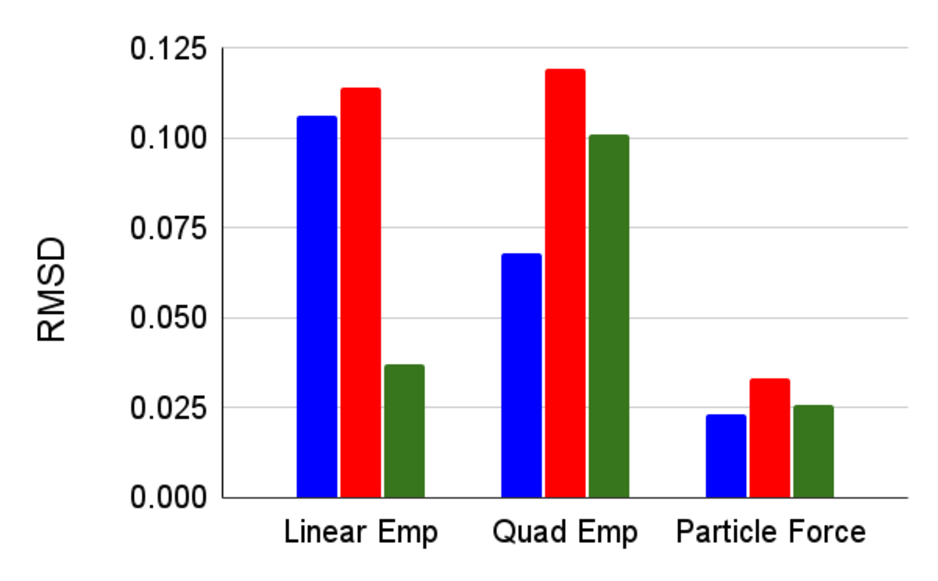}\put(-130,105){(b)}
    \includegraphics[scale=0.375]{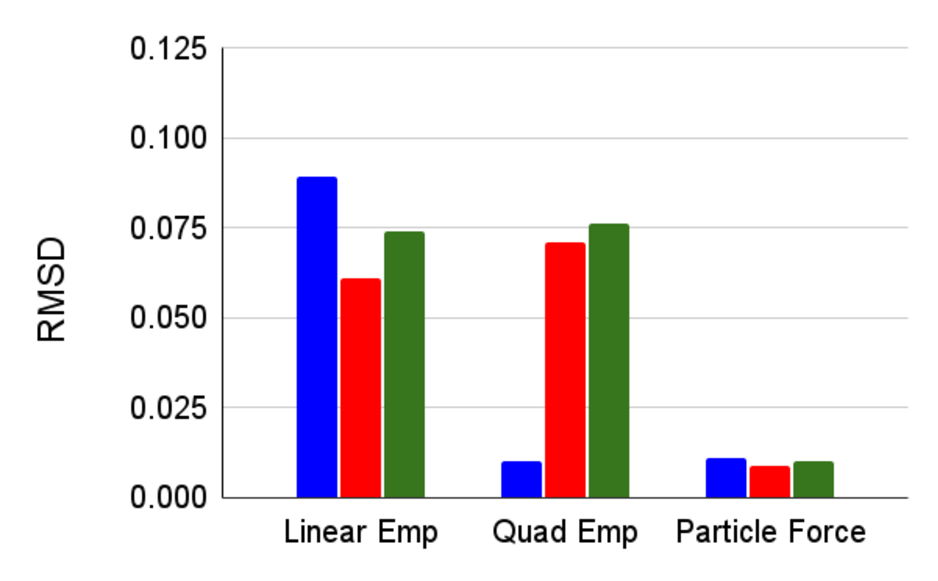}\put(-130,105){(c)}
    \caption{The root mean square deviation (RMSD) values for light (blue), medium (red) and heavy (green) species using three different segregation models in ternary mixtures of density ratio $\rho_H : \rho_M : \rho_L = 3:2:1$. Three different mixture compositions of (a) $f^T_L = 70\%, f^T_M = 20\%, f^T_H = 10\%$, (b) $f^T_L = 10\%, f^T_M = 70\%, f^T_H = 20\%$, and (c) $f^T_L = 20\%, f^T_M = 10\%, f^T_H = 70\%$ are considered. }
    \vspace{1.7cm}
    \label{fig:ternary-rmsd}
\end{figure*}
These results confirm that both the quadratic empirical model and the particle force-based model are capable of accurately predicting segregation for equal composition ternary mixture.
In order to explore the applicability of the empirical model over a wide range of mixture compositions, we select mixtures with $70\%$ composition of a particular species and $20\%$ and $10\%$ of the other two species. We obtain the theoretical predictions for each of these mixtures and compare them with their DEM simulations.
\begin{figure*}[!htb]
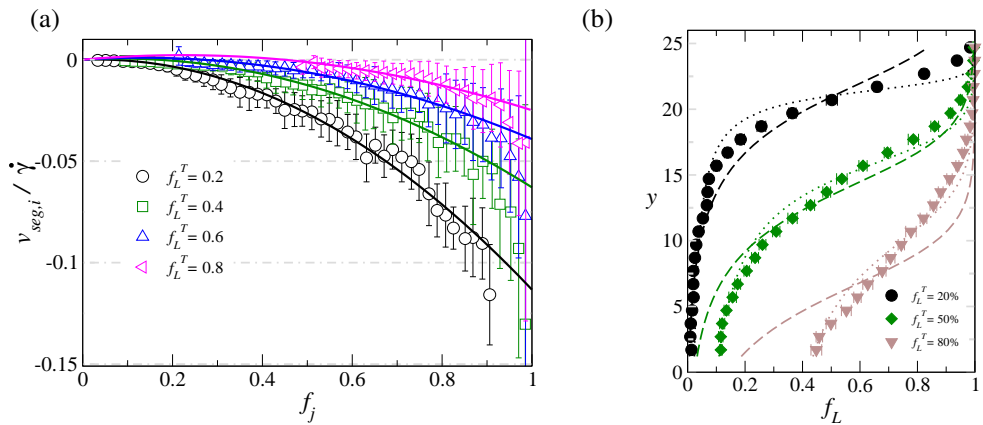

    \centering
  \includegraphics[scale=0.365]{Figures/Fig14a.eps}\put(-190,150){(a)} \quad \quad \quad \quad
     \includegraphics[scale=0.35]{Figures/Fig14b.eps}\put(-150,150){(b)}\hfill
    \caption{(a) Shear scaled segregation velocity of heavy species for different composition mixture. 
    Symbols represent DEM data, and solid lines represent the quadratic empirical fits of form $v_{seg,i}/\dot{\gamma}= d f_j (A_{ij} + B_{ij}) f_j$. (b) Steady-state concentration profiles of light species in a binary mixture of density ratio $\rho = 1.5$.
    Symbols represent DEM data. Lines represent quadratic empirical model predictions. Dashed lines correspond to the empirical parameters estimated by averaging data over a wide range of compositions while dotted lines correspond to empirical parameters estimated for individual mixture compositions of light species.}
    \label{fig:Conc-depend-lowdensity}
\end{figure*}

Figure~\ref{fig:ternary-conc-diffComp}a shows the results for the mixture with $70\%$ of light species i.e., $f^T_L = 0.70$ with $f^T_M = 0.20$ \& $f^T_H = 0.10$. Theoretical predictions using the linear empirical segregation model (dashed lines) match well with the DEM data for all three species. In contrast to the equal composition mixtures case, the quadratic empirical segregation model predictions (dotted lines) deviate strongly from the DEM results and particle force-based segregation model predictions. 
In this case, the quadratic approach fails to capture the DEM results throughout the entire layer height. Figure~\ref{fig:ternary-conc-diffComp}b shows predictions for the mixture with $70\%$ medium species composition; the heavy and light species composition being $20\%$ and $10\%$, respectively. For this mixture, the predictions using the quadratic empirical model as well as the linear empirical model differ substantially from the DEM data for both light (blue) and medium (red) species. On the other hand, the predictions based on the particle force model exhibit very good match with the DEM data. Figure~\ref{fig:ternary-conc-diffComp}c shows the results for a ternary mixture where the composition of heavy species is $70\%$ and that of the light species is $20\%$. Predictions using the quadratic approach for the light species are found to be in good agreement with DEM data. Differences for the medium-density and high-density species are noticeable. The linear empirical segregation model (dashed line) shows significant deviation for all three species. Theoretical predictions obtained using the particle force-based model exhibit excellent agreement with the DEM data for all three species for this case as well.
These observations can be confirmed by comparing the RMSD values of the three models for each of the mixtures. Figure~\ref{fig:ternary-rmsd} shows the RMSD value for the light (blue), medium (red) and heavy (green) species for each segregation model. In all the cases, the RMSD value for the particle force-based theory is much smaller compared to both the empirical models. This confirms that the particle force-based segregation model is able to capture the segregation of various kinds of mixtures. In contrast, the linear and quadratic empirical models seem to work well in some cases and fail in some other situations.

In order to understand the inability of the quadratic model to predict concentration in the mixtures rich in either in light or in medium species, we plot the shear rate scaled segregation velocity of the heavy species for four different compositions of binary mixtures in figure~\ref{fig:Conc-depend-lowdensity}a. Data have been shown for a binary mixture of density ratio $\rho = 1.5$ flowing over an inclination angle of $\theta = 25\degree$. Different symbols correspond to different mixture compositions. Evidently, the variation of the shear rate scaled segregation velocity is different for different compositions of the mixture. For example, the value of $v_{seg,i}/\dot{\gamma}$ for a mixture composition of $f^T_L = 0.8$ is less than one-fourth of the value for a composition mixture of $f^T_L = 0.2$. The magnitude of $v_{seg,i}/\dot{\gamma}$ of heavy species decreases with an increase in the composition of light species in the mixture. Hence the empirical parameters obtained for \textcolor{black}{different mixture} compositions are expected to show significant differences.
\begin{table}[h]
\centering
\begin{tabular}{cccccc}
\hline
$f^T_L$ & 0.20 & 0.40 & 0.50 & 0.60 & 0.80 \\
\hline
$A_{HL}$ & 0.013 & 0.040 & 0.037 & 0.032 & 0.037 \\
$B_{HL}$ & -0.132 & -0.123 & -0.101 & -0.084 & -0.070 \\
\hline
\end{tabular}
\caption{Quadratic empirical parameters for heavy species for five different composition mixtures of density ratio $\rho = 1.5$.}
\label{Table:Quad_parameters_DiffComp}
\end{table}

Table~\ref{Table:Quad_parameters_DiffComp} shows the parameters for each composition obtained by fitting the quadratic model to the data shown in figure~\ref{fig:Conc-depend-lowdensity}a. Solid lines in figure~\ref{fig:Conc-depend-lowdensity}a show the fitted quadratic curve using the values reported in Table~\ref{Table:Quad_parameters_DiffComp}.
Figure \ref{fig:Conc-depend-lowdensity}b shows the predictions obtained using the quadratic empirical parameters for the respective composition of the mixture using dotted lines. The predictions match accurately with DEM data for all three compositions. However, employing the averaged empirical parameters across all the compositions, the predictions (shown by dashed lines) deviate significantly from the DEM data. 
The RMSD values for the dashed lines (using the same parameters for all compositions) are found to be $0.06$, $0.06$, \& $0.10$ for  $f_L^T=20\%, 50\%$, and $80\%$ mixtures, respectively. However, using the values corresponding to the mixture composition, the RMSD values reduce to $0.05$, $0.04$, \& $0.03$, respectively. 
\textcolor{black}{Our analysis confirms that the large RMSD values for $f^T_L > 0.5$ in the quadratic empirical model case shown in figure~\ref{fig:RMSD} can be attributed to this concentration dependence of the parameters. For example, using the revised parameters for $\rho=3.0$ and $f_L^T=80\%$ mixture (figure~\ref{fig:RMSD}c) reduces the RMSD value to $0.02$ from $0.08$.}
Thus we conclude that the quadratic empirical parameters indeed depend on the total composition of the species in the mixture. This is in contrast to the observations reported by \cite{jones2018asymmetric} who reported that the inlet composition in heap flow does not affect the empirical segregation velocity relation for quadratic segregation models. A similar comparison for the linear empirical model, however, does not show the dependency of the fitted parameters on mixture composition. This is in agreement with \cite{xiao2016modelling} who showed that the RMSD values for the predictions do not change significantly when using the linear empirical parameters obtained for individual composition.

\section{Conclusion}
\label{sec:conclusion}
In this study, we have investigated the suitability of empirical segregation models for predicting segregation in periodic chute flows. This simple system allows us to explore the effect of the inclination angle, different initial configurations, and the duration of measurement on the segregation parameters.
We explore the two existing empirical approaches by accounting for or ignoring diffusional contribution in the computation of segregation velocity $v_{seg,i}$ from DEM simulations. While the parameters are estimated using transient data in the case of the linear empirical model, the quadratic model uses steady-state data for estimating the parameters. We find that the time period of measurement of $v_{seg,i}$ affects the values of the empirical parameters estimated in the case of the linear empirical model. \textcolor{black}{In addition, the initial configuration also seems to have a significant effect on the estimated values of these parameters.} The appropriate values of empirical parameters in the case of the linear empirical model can be obtained by measuring percolation velocity for approximately twice the segregation time scale in a system starting with mixed configuration. \textcolor{black}{The other configurations should be avoided for determining the linear empirical model parameters.} The quadratic model parameters, however, are found to be less sensitive to the initial configurations used. In addition, the time scale of observation does not affect these values as well since steady-state data are used. 

Using the appropriate values of empirical parameters for both the empirical models, we predict the concentration profiles for mixtures of different compositions and density ratios. \textcolor{black}{In addition, we also predict the concentration profile from the particle force-based segregation model.} We compute RMSD values to quantify the deviation of the predicted profiles with the corresponding DEM simulation data. 
\textcolor{black}{A comparison of the RMSD values shows that the empirical model RMSD values are significantly larger than those of the particle force-based model predictions.}
The linear empirical model has a larger RMS deviation for mixtures rich in heavy species. The quadratic empirical model predictions have larger RMS deviation in the case of mixture rich in light species. The deviation in the quadratic model case arises due to the dependence of empirical parameters on mixture composition and can be substantially reduced by using parameters corresponding to the specific composition. 

We also compare the predictions of the empirical segregation models and particle force-based theory with DEM simulations for ternary mixtures. While the linear empirical as well as particle force-based models have been used to predict segregation in multi-component mixtures, the quadratic model predictions have been restricted to binary mixtures only. We extended the methodology to predict segregation in the case of ternary mixtures using quadratic empirical model. A comparison of the predictions from these three segregation models for ternary mixtures of different compositions reconfirms that the particle force-based theory is able to predict the simulation data in much better agreement compared to the empirical models. 
Owing to the fact that the segregation model in the particle force-based approach is obtained from the forces acting on the particles, \textcolor{black}{the species segregation velocity in this approach accounts for the effect of the inclination angle and viscosity dependence on local mixture composition. Such effects are not accounted in the empirical approach.}

A similar comparison of empirical and particle force-based segregation models also needs to be done for mixtures having different sizes.
It remains to be seen whether the dependencies of the segregation parameters on the measurement time, initial configuration, and inclination angle are observable for empirical models in the case of size segregation. In addition, mixtures differing in both size and density also need to be considered. While empirical models for such mixtures have been proposed, a particle force-based theory for the combined effect of size and density is still lacking and will be reported in the near future. Finally, the particle force-based theory also needs to be explored in few other flow configurations to establish its applicability in experimentally realizable systems.

\backsection[Funding]{AT gratefully acknowledges the financial support provided by the Indian Institute of Technology Kanpur via the initiation grant IITK/CHE/20130338. AT and SK gratefully acknowledge the funding support provided to SK by the Prime Minister's Research Fellowship (Government of India) grant.}
\backsection[Declaration of Interest]{The authors report no conflict of interest.}
\backsection[Author ORCIDs]{
\newline
\\
Soniya Kumawat  \url{https://orcid.org/0000-0003-3314-9875};\\
\\
Vishnu Kumar Sahu \url{https://orcid.org/0000-0002-7768-0505}; \\
\\
Anurag Tripathi \url{ https://orcid.org/0000-0001-9945-3197}.}
\bibliographystyle{elsarticle-harv} 
\bibliography{seg-bib}
 \newpage
 \appendix
 \beginsupplement
 \setcounter{equation}{0}
 \setcounter{page}{1}
\section{Effect of diffusivity values on segregation predictions}

\textcolor{black}{\cite{fan2014modelling} and \cite{GAO2021} considered both spatially varying diffusivity values as well as a constant diffusivity value in the prediction of binary size segregation using linear empirical model. The authors report that the spatial variation in the diffusivity has a negligible effect on the predictions of the continuum model in the case of quasi-2D heap flows.} 
Table~\ref{table:Diffusivity-value} shows that while the mean diffusivity values for a given inclination do not differ a lot from each other due to change in mixture composition, the effect of inclination on the mean diffusivity is very significant and hence needs to be accounted for if one wishes to use a constant diffusivity.
\vspace{1cm}
\begin{table}[!htb]
    \centering
    \begin{tabular}{cccccc}
        \toprule
        $f^T_L$ & 0.20 & 0.40 & 0.50 & 0.60 & 0.80 \\
        \midrule
        $\theta = 25\degree$ & 0.0121 & 0.0114 & 0.0113 & 0.0114 & 0.0127 \\
        $\theta = 29\degree$ & 0.0361 & 0.0325 & 0.0321 & 0.0322 & 0.0345 \\
        \bottomrule
    \end{tabular}
    \caption{Mean diffusivity ($D_{avg}$) values for five different binary mixture compositions at two different inclination angles $\theta = 25\degree$ and $29\degree$.}
    \label{table:Diffusivity-value}
\end{table}
\vspace{1cm}

\begin{figure}
    \centering
   \quad \quad   \quad \quad   \quad  \includegraphics[scale=0.4]{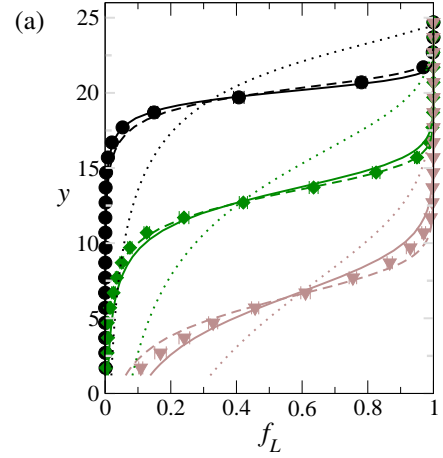}\put(-160,160){(a)}
    \hfill
    \vspace{2cm}
     \includegraphics[scale=0.4]{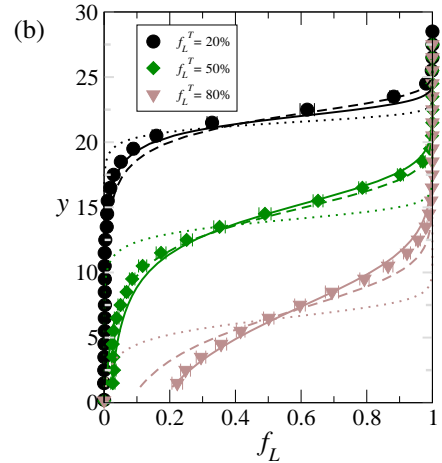}\put(-160,160){(b)}
    \caption{Comparison of theoretical predictions using a constant diffusivity and a spatially varying diffusivity $D = b \dot \gamma d^2_{mix}$ for a binary mixture of density ratio $\rho = 3.0$ for (a) $\theta = 25\degree$ and (b) $\theta = 29\degree$. Symbols represent DEM data, solid lines represent the predictions using spatially varying diffusivity $D = b \dot \gamma d^2_{mix}$. Dashed lines represent the predictions using constant diffusivity $D$ corresponding to inclination angle $\theta$ for $50\% - 50\%$ binary mixture. Dotted lines represent the predictions using a constant value of diffusivity $D$ taken for $\theta=29^\circ$ in (a) and $\theta=25^\circ$ in (b).}
    \label{fig:diff_supfig}
\end{figure}

Figure~\ref{fig:diff_supfig} shows the comparison of theoretical predictions using particle force-based theory using both constant as well as spatially varying diffusivity across the layer. Data are shown for three different binary mixtures having light species composition $f^T_L = 0.20$ (black), $f^T_L = 0.50$ (green), and $f^T_L = 0.80$ (brown) with density ratio $\rho = 3.0$. 
Figure~\ref{fig:diff_supfig}a shows the predictions at inclination angle $\theta = 25\degree$ while figure~\ref{fig:diff_supfig}b shows the predictions at inclination angle $\theta = 29\degree$. Solid lines represent the theoretical predictions accounting for the spatial variation of the diffusivity using the relation $D = b \dot \gamma d^2_{mix}$. The dashed lines represent the predictions using a constant value of average diffusivity $D_{avg}$ corresponding to the inclination angle for $50\% - 50\%$ binary mixture. Both solid and dashed lines accurately match with the DEM data (symbols) for all three composition mixtures at two different inclination angles. 

Dotted lines in figure~\ref{fig:diff_supfig}a correspond to predictions obtained using $D_{avg}$ for $\theta=29\degree$ instead of $25\degree$. The discrepancy from the DEM simulation data (symbols) is evident. This mismatch confirms that using the average value of diffusivity at one angle for predicting segregation of mixtures flowing at another angle gives highly inaccurate results. 
Similarly, using data for $D_{avg}$ corresponding to $\theta=25\degree$ to predict segregation for $\theta=29\degree$ case also leads to highly inaccurate predictions (figure~\ref{fig:diff_supfig}b). The difference is attributed to the fact that the diffusivity values at $\theta=29\degree$ are nearly three times larger than $\theta=25\degree$ case. 
This indicates that accounting for the variation of the diffusivity with the shear rate using the relation $D = b \dot \gamma d^2_{mix}$ should be preferred as it captures the influence of the inclination angle on the diffusivity as well.

\end{document}